\documentclass[prd,preprint,tightenlines,showpacs,preprintnumbers,nofootinbib,eqsecnum,superscriptaddress]{revtex4-1}

 \usepackage[dvips,final]{graphicx}
  \usepackage{amssymb}
   \usepackage{amsmath}
    \usepackage{amsfonts}
     \usepackage{epsfig}
      \usepackage{bm}
      \usepackage[dvipsnames,usenames]{color}
      \usepackage{comment}
       \usepackage[utf8]{inputenc}
        \usepackage{soul}

\usepackage[section]{placeins}

\usepackage{sidecap}
\usepackage{multirow}
\usepackage{booktabs}
\usepackage{array}
\usepackage{tabularx}
\usepackage{xcolor}
\usepackage{pstricks}
\def\frac#1#2{{\begingroup #1\endgroup\over #2}}
\def\Pom{\mathrm{I\!P}}

\newcommand{\GeV}{\textrm{GeV}}

\newcommand{\half}{{1\over 2}}

\begin{document}
\title{Probing gluon GTMDs of the proton in deep inelastic diffractive dijet production at HERA}

\author{Barbara Linek}
\email{barbarali@dokt.ur.edu.pl}
\affiliation{University of Rzesz\'ow, ul. Pigonia 1, PL-35-959 Rzesz\'ow, Poland}
\author{Marta Łuszczak}
\email{mluszczak@ur.edu.pl}
\affiliation{University of Rzesz\'ow, ul. Pigonia 1, PL-35-959 Rzesz\'ow, Poland}
\author{Wolfgang Sch\"afer}
\email{Wolfgang.Schafer@ifj.edu.pl} 
\affiliation{Institute of Nuclear
Physics, Polish Academy of Sciences, ul. Radzikowskiego 152, PL-31-342 
Krak{\'o}w, Poland}
\author{Antoni Szczurek}
\email{antoni.szczurek@ifj.edu.pl}
\affiliation{Institute of Nuclear Physics, Polish Academy of Sciences, 
ul. Radzikowskiego 152, PL-31-342 Krak{\'o}w, Poland}

\begin{abstract}
We calculate several differential distributions for diffractive dijets production in $e p \to e'  {\rm jet \, jet} p$ in the pQCD dipole approach using off diagonal unintegrated gluon distributions (GTMDs). Different models from the literature are used. We concentrate on the contribution from exclusive  $q \bar q$ dijets. 

Results of our calculations are compared to H1 and ZEUS data, including specific experimental cuts in our calculations. In general, except of one GTMD, our results are below the HERA data. The considered mechanism is expected to gives a sizeable contribution to the ZEUS data, while it is negligible in the kinematics of the H1 measurement.

This is in contrast to recent results from the literature where the normalization was adjusted to some selected distributions of H1 collaboration and no agreement with other observables was checked.

The ZEUS data provide stricter limitations on the GTMDs than the H1 data. We conclude, based on comparison to different observables, that the calculated cross sections are only a small fraction of the measured ones which contain probably also processes with pomeron remnant. Alternatively the experimental data could be explained by inclusion of $q \bar q g$ component. 

We present also azimuthal correlations between the sum and the difference of dijet transverse momenta.
The cuts on transverse momenta of jets generate corresponding azimuthal correlations which can be misidentified as due to elliptic gluon distributions.

\end{abstract}
\maketitle

\section{Introduction}
The $e p$ scattering at the HERA collider was proven to be an ideal tool for studying proton structure and in particular gluon distributions in the proton in the region of very small Bjorken-$x$ or small gluon momentum fraction \cite{Newman:2013ada}. The dijet production is a particularly valuable final state in this context. 
Here we concentrate on the exclusive, or diffractive, dijet production in the $e p \to e jj p$ reaction, with the final state proton in its ground state as measured at HERA
by the H1 \cite{H1:2012} and ZEUS \cite{ZEUS:2016} collaborations. 
Experimentally, the exclusivity requirement on the final state is very demanding. Often rather a semiexclusive reaction $e p \to j j  x p $ or even $ep \to jj x p^*$ is measured, where $x$ denotes additional hadronic activity in the diffractive system, besides the dijet, whereas $p^*$ stands for a possible excitation of the proton with particle production in its fragmentation region. 

Diffractive processes, which fulfill the additional requirement of a large rapidity gap between proton and the diffractive final state, are a testbed on ideas for the QCD Pomeron. These processes are conveniently described in the color dipole approach, where the leading order process for diffractive masses $M^2 \sim Q^2$
is the diffractive excitation of the $q \bar q$ state of the virtual photon \cite{Nikolaev:1991et}. 
There are several related approaches to that problem, see for example the reviews \cite{Wusthoff:1999cr,Ashery:2006zw,Frankfurt:2022jns}. One of them relies on the Pomeron having partonic structure,  and is based on collinear factorization involving diffractive parton distributions \cite{Watt:2007ee}. This approach strictly applies to inclusive diffraction. Calculations have been performed to high orders (NNLO) in the perturbative expansion \cite{Britzger:2018zvv}. The photoproduction limit, which has applications to ultraperipheral heavy ion collisions has been addressed in this framework in \cite{Guzey:2018dlm}.

In this work, we adopt the formalism which starts from the color dipole approach, but is formulated in momentum space. Here the information contained in the impact parameter space dipole amplitude is transferred to off-forward transverse momentum dependent gluon distributions, commonly referred to as GTMDs (generalized transverse momentum dependent distributions) in the literature. For reviews which explore the relation to the gluon Wigner function, see Refs. \cite{Pasechnik:2023mdd,Boussarie:2023izj}. 
At large jet momenta the forward diffractive amplitude probes directly the unintegrated gluon distribution of the target \cite{Nikolaev:1994cd, Nikolaev:2000sh}. This approach is appropriate in the small-$x$ or high-energy limit, and takes into account only the transverse momentum transfer. The longitudinal momentum transfer, and the so-called skewedness is taken into account in a collinear factorization formalism based on generalized parton distributions used in Ref. \cite{Braun:2005rg}. This work also includes $q \bar q$ exchanges in the $t$-channel, which are relevant at not too large rapidity gap.
In Ref. \cite{Linek:2023kga} we discussed different models of GTMDs in application to the $p  A \to c \bar c p A$ process. In this case no data is available so far and the measurement is challenging. Here we will use essentially the same formalism to the case in which both H1 and ZEUS data are available. We intend to compare the results of the formalism and different GTMDs to the existing data.
Other recent calculations of diffractive dijet production in the color dipole or GTMD approaches are in Refs. \cite{Altinoluk:2015dpi,Salazar:2019ncp,Mantysaari:2020lhf,PhysRevD.100.074020,Hagiwara:2016kam,Hagiwara:2017fye,ReinkePelicer:2018gyh,Boer:2021upt,Boer:2023mip}.
Some of these papers concentrate on photoproduction or heavy quarks only. We have some overlap with the work \cite{PhysRevD.100.074020}, who use the Golec-Biernat--W\"usthoff parametrization \cite{Golec-Biernat:1998zce} of the dipole amplitude. For the corresponding gluon distribution, our results and conclusions agree with theirs.
We also use the GTMDs proposed/fitted in Refs.\cite{Boer:2021upt,Boer:2023mip}, however we differ in our conclusions from these works.
\section{Formalism}
\label{sec:Formalism}
In Fig.1 we show the four pQCD diagrams contributing to the exclusive diffractive production of dijets.
 \begin{figure}[h]
  \centering
  \includegraphics[width=.242\textwidth]{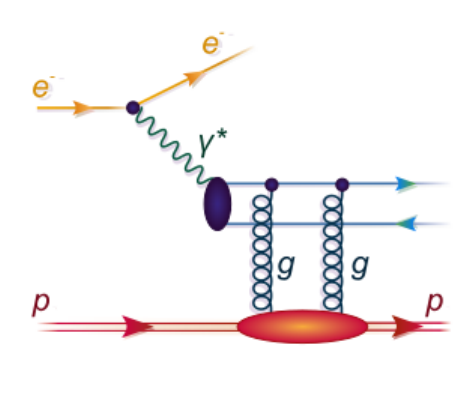}
  \includegraphics[width=.242\textwidth]{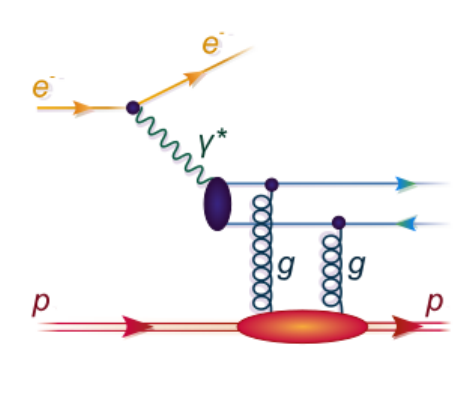}
  \includegraphics[width=.242\textwidth]{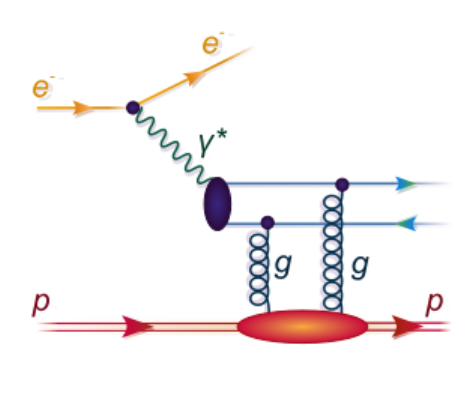}
  \includegraphics[width=.242\textwidth]{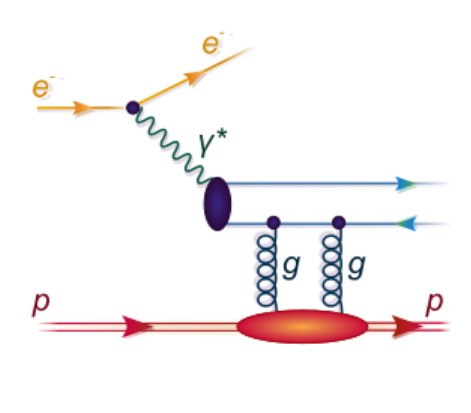}  
  \caption{Feynman diagrams for the diffractive production of dijets in electron-proton collisions, discussed in the present paper.}
  \label{fig:diagrams}
\end{figure}

\subsection{Kinematic variables}
In this paper we will use the standard kinematic variables: 
for incoming and outgoing lepton four-momenta $k,k'$, we have the photon four-momentum $q = k - k'$, with $q^2 = - Q^2 <0$. Let $P$ be the four-momentum of the incoming proton, then the $ep$ cm-energy squared is
$ s = (k+P)^2 \sim 2 k\cdot P$
and the so-called inelasticity $y$ is defined as
\begin{eqnarray}
    y = \frac{ q \cdot P}{k \cdot P} \, ,
\end{eqnarray}
so that the square of cms energy of the $\gamma^* p$ system $ W^2_{\gamma p} = (q + P)^2$ is calculated from 
\begin{eqnarray}
W_{\gamma p}^{2} = ys-Q^2,
\end{eqnarray}
where we neglected the proton mass.
The standard Bjorken variable $x_{Bj}$
\begin{eqnarray}
x_{Bj}=\frac {Q^2}{2 P \cdot q} = \frac{Q^{2}}{Q^{2}+W_{\gamma p}^{2}-m_{p}^{2}} \, .
\end{eqnarray}
The diffractive jets carry momentum fractions $z, 1-z$ of the photon and momenta $\vec p_{\perp1}, \vec p_{\perp 2}$ transverse to the photon-proton collision axis.
The square of the dijet invariant mass is
\begin{eqnarray}
M_{jj}^{2} = \frac{p_{\perp 1}^{2}+m_{q}^2}{z} + \frac{p_{\perp 2}^{2}+m_{q}^{2}}{1-z}-\Delta_{\perp}^{2},
\end{eqnarray}
where $\vec \Delta_\perp = \vec p_{\perp1} + \vec p_{\perp2}$ is the transverse momentum of the dijet system, which for the exclusive limit of interest here, equals (negative of) the momentum transfer to the proton. 
We also use the standard diffractive variables
\begin{eqnarray}
\beta = \frac{Q^{2}}{Q^{2}+M_{jj}^{2}},
\end{eqnarray}
which is the fraction of the Pomeron momentum carried by the struck quark. 
The fraction of the proton momentum carried by the pomeron is
\begin{eqnarray}
x_{\Pom} = \frac{x_{Bj}}{\beta}.
\end{eqnarray}
It is also interesting to calculate the azimuthal angle correlations between the sum and the difference of the jets transverse momenta
\begin{eqnarray}
\cos \phi_{\vec P_\perp \vec \Delta_\perp}  = \frac{\vec P_\perp \cdot \vec \Delta_\perp}{P_\perp \Delta_\perp} \, ,
\end{eqnarray}
where 
\begin{eqnarray}
    \vec P_\perp = \half( \vec p_{\perp 1} - \vec p_{\perp 2} )\, , \qquad 
    \vec \Delta_\perp = \vec p_{\perp 1} + \vec p_{\perp 2} \, . 
\end{eqnarray}
We will also use jet rapidities $\eta_{1}$ and $\eta_{2}$ which are important only when experimental cuts are imposed (see Table~\ref{tab:H1andZEUScuts}).

\subsection{Cross section}

In order to compare to HERA data with appropriate experimental cuts, we want to perform our calculations at the $ep$ level. 
The cross section can be written in terms of photon fluxes and $\gamma^* p$ cross sections as
\begin{eqnarray}
\frac{d\sigma^{ep}}{dy dQ^{2} d\xi} &=&\frac{\alpha_{em}}{\pi y Q^{2}} \Big[ \Big(1-y+\frac{y^{2}}{2} \Big) \frac{d \sigma_{T}^{\gamma^{*}p}}{d \xi}+(1-y) \frac{d\sigma_{L}^{\gamma^{*}p}}{d\xi}  \Big].
\end{eqnarray}
Here $d \xi = dz d^2\vec P_\perp d^2 \vec \Delta_\perp$, and $T,L$ stand for the transverse and longitudinal photons.
We have neglected interferences between photon polarizations, which vanish if one averages over the angle between electron scattering plane and hadronic plane.
The $\gamma^* p \to q \bar q p$ cross sections read 
\begin{eqnarray}
\frac{d\sigma^{\gamma^{*}p}_{T}}{dzd^{2}\vec P_{\perp}d^{2} \vec \Delta_{\perp}}&=&2N_{c}\alpha_{em}\sum_{f}e_{f}^{2}\int d^{2}\vec k_{\perp}\int d^{2}\vec k^{~'}_{\perp}T \big(Y,\vec k_{\perp},\vec \Delta_{\perp}\big) T\big(Y,\vec k^{~'}_{\perp},\vec \Delta_{\perp} \big) \nonumber\\ 
 &\times& \Bigg\{ \Big(z^2 + (1-z)^2 \Big)\left[\frac{(\vec P_{\perp}-\vec k_{\perp})}{(\vec P_{\perp}-\vec k_{\perp})^{2} +\epsilon^2}-\frac{\vec P_{\perp}}{P^{2}_{\perp}+\epsilon^2}\right] \cdot \left[\frac{(\vec P_{\perp}-\vec k^{~'}_{\perp})}{(\vec P_{\perp}-\vec k^{~'}_{\perp})^{2} + \epsilon^2}-\frac{\vec P_{\perp}}{P^{2}_{\perp}+\epsilon^2}\right] \nonumber \\
 &+& m_f^2 \left[\frac{1}{(\vec P_{\perp}-\vec k_{\perp})^{2}
 +\epsilon^2}-\frac{1}{P^{2}_{\perp}+\epsilon^2}\right] \cdot \left[\frac{1}{(\vec P_{\perp}-\vec k^{~'}_{\perp})^{2}+\epsilon^2}-\frac{1}{P^{2}_{\perp}+\epsilon^2}\right] \Bigg\},
\end{eqnarray}
\begin{eqnarray}
\frac{d\sigma^{\gamma^{*}p}_{L}}{dzd^{2}\vec P_{\perp}d^{2} \vec\Delta_{\perp}}&=&2N_{c}\alpha_{em} 4Q^2 z^2 (1-z)^2
\times \sum_{f}e_{f}^{2}\int d^{2}\vec k_{\perp}\int d^{2}\vec k^{~'}_{\perp}T \big(Y,\vec k_{\perp},\vec \Delta_{\perp}\big) T\big(Y,\vec k^{~'}_{\perp},\vec \Delta_{\perp} \big) \nonumber\\ 
 &\times& \left[\frac{1}{(\vec P_{\perp}-\vec k_{\perp})^{2} + \epsilon^2}-\frac{1}{P^{2}_{\perp}+\epsilon^2 }\right] \cdot \left[\frac{1}{(\vec P_{\perp}-\vec k^{~'}_{\perp})^{2} + \epsilon^2}-\frac{1}{P^{2}_{\perp}+\epsilon^2}\right],
\end{eqnarray}
with
\begin{eqnarray}
\epsilon^{2}=z(1-z)Q^{2}+m_f^2.
\end{eqnarray}
Notice, that as e.g. in Refs. \cite{Boer:2021upt,Boer:2023mip}, the transverse momentum $\vec \Delta_\perp$ enters only into the GTMD, and the so-called impact factor that describes the $\gamma^* \to q \bar q$ transition and coupling of quarks to gluons, is taken in the forward direction. We do not discuss the
subtleties related to this approximation (see for example Ref. \cite{Linek:2023kga}), which would become crucial only in the discussion of more subtle angular correlations. 
Our formulae here differ e.g. from those in ~\cite{Boer:2021upt}, where quark masses were neglected. The effect of mass is important for the $c\bar{c}$ dijets, as will be discussed in the results section.

The unintegrated gluon density matrix $f$ encodes the same information as the generalized transverse momentum distribution (GTMD) of gluons in the proton target. It was proven that in practical numerical applications $f$ and $T$ agree up to the factor $1/2$ (for details see Ref. ~\cite{Linek:2023kga})
\begin{eqnarray}
    f\Big(Y,\frac{\vec \Delta_\perp}{2} + \vec k_\perp, \frac{\vec \Delta_\perp}{2} - \vec k_\perp\Big) \to - \half \, T(Y,\vec k_\perp, \vec \Delta_\perp) \, .
\end{eqnarray}
\subsection{GTMDs}
The mentioned GTMD is a representation of the diffraction amplitude in momentum space often used in the literature (see e.g. Refs. \cite{Hagiwara:2016kam,Hagiwara:2017fye,ReinkePelicer:2018gyh,Pasechnik:2023mdd}) which is the Fourier transform of the dipole amplitude 
\begin{eqnarray}
    T(Y,\vec k_\perp, \vec \Delta_\perp)
    = \int \frac{d^2 \vec b_\perp}{(2 \pi)^2} \frac{d^2 \vec r_\perp}{(2 \pi)^2} \, e^{-i \vec \Delta_\perp \cdot \vec b_\perp} \, e^{- i \vec k_\perp \cdot \vec r_\perp} \, 
     N(Y,\vec r_\perp, \vec b_\perp) \, .
     \label{eq:T_definition}
\end{eqnarray}
We decided to choose normalization that is consistent with Ref.~\cite{ReinkePelicer:2018gyh}. However, such a Fourier transform does not converge to zero and requires regularization, which is often a Gaussian cutoff function
\begin{eqnarray}
    T(Y,\vec k_\perp, \vec \Delta_\perp) = \int \frac{d^2 \vec b_\perp}{(2 \pi)^2} \frac{d^2 \vec r_\perp}{(2 \pi)^2}  
    e^{-i \vec \Delta_\perp \cdot \vec b_\perp} e^{-i \vec k_\perp \cdot \vec r_\perp} \, N(Y,\vec r_\perp, \vec b_\perp) \, e^{-  \varepsilon r_\perp^2}\, .
     \label{eq:reg}
\end{eqnarray}
The value of the $\varepsilon$ parameter has a significant impact on the distribution of the obtained cross-sections, which was shown in Ref. \cite{Linek:2023kga}. For the purposes of this analysis, we decided to assume $\varepsilon = (0.5 \, \rm{fm})^{-2}$, similarly to Refs.~\cite{ReinkePelicer:2018gyh,Boer:2021upt,Linek:2023kga}. The Fourier expansion of the dipole amplitude allows to distinguish an isotropic contribution and an elliptical term depending on the orientation of the dipole,
\begin{eqnarray}
    N(Y,\vec r_\perp, \vec b_\perp) = N_0(Y, r_\perp, b_\perp) + 2 \, \cos (2 \phi_{br}) \, N_\epsilon(Y,r_\perp, b_\perp) + \dots \,,
\end{eqnarray}
however we only introduce the isotropic part of the dipole amplitude in the present analysis.

We consider five different GTMD models. Two of them are parameterizations of off-forward gluon density matrices for diagonal unintegrated gluon distribution compatible with the Golec-Biernat-W\"usthoff (GBW model) \cite{Golec-Biernat:1998zce} and Moriggi-Paccini-Machado (MPM model) \cite{Moriggi:2020zbv} for which we use the diffractive slope $B=4$ $\GeV^{-2}$,

\begin{eqnarray}
     f\Big(Y,\frac{\vec \Delta_\perp}{2} + \vec k_\perp, \frac{\vec \Delta_\perp}{2} - \vec k_\perp\Big) = \frac{\alpha_s}{4 \pi N_c} \, \frac{{\cal F}(x_\Pom, \vec k_\perp, -\vec k_\perp)}{k_\perp^4} \, \exp\Big[ - \half B \vec \Delta^2 \Big] \, . 
     \label{eq:model_f}
\end{eqnarray}
We also included models that are the Fourier transforms of the dipole amplitude described by equation (\ref{eq:reg}). We use also the bSat model of Kowalski and Teaney \cite{Kowalski:2003hm} (KT model) with parameters from Table I fit 3 of \cite{Kowalski:2003hm} and two models based on the McLerran-Venugopalan approach ~\cite{McLerran:1993ka} - proposed by Iancu-Rezaeian ~\cite{Iancu:2017fzn} (MV-IR model) and the Boer-Setyadi ~\cite{Boer:2021upt} (MV-BS model) fitted to the H1 experimental data. Both of these models are independent of the $x_{\Pom}$, therefore we modified the MV-IR model using $\lambda = 0.277$ as 
\begin{eqnarray}
    T_{\rm MV-IR}^{\rm mod}(Y,\vec k_\perp, \vec \Delta_\perp) = T_{\rm MV-IR}(\vec k_\perp, \vec \Delta_\perp) \, e^{\lambda Y}, \quad \, Y = \ln\Big[\frac{0.01}{x_\Pom}\Big] \,.
\end{eqnarray}
It should be noted that more realistic extensions of the MV-IR model have been proposed in the literature, see e.g. Ref. \cite{Mantysaari:2020lhf}. 
Adapting the MV-BS model according to ~\cite{Boer:2021upt} we introduce $\chi=1.25$ (see below),

\begin{eqnarray}
N_0(r_\perp, b_\perp)=-\frac{1}{4}r_{\perp}^{2}\chi Q_{s}^{2}(b_{\perp})\ln \Big[\frac{1}{r_{\perp}^{2}\lambda^{2}} +e \Big],
\end{eqnarray}
with
\begin{eqnarray}
   Q_{s}^{2}(b_{\perp})=\frac{4\pi\alpha_{s}C_{F}}{N_{c}} \exp \Big[ \frac{-b_{\perp}^{2}}{2R_{p}^{2}} \Big]\,.
\end{eqnarray}

In principle, $\chi$ can be treated as a free parameter to be adjusting to experimental data.

\section{Numerical results}
\label{sec:distributions}
Here we will present our results for both H1 and ZEUS cuts. The details of the cuts are described in \cite{H1:2012} and \cite{ZEUS:2016} papers and summarized in Table~\ref{tab:H1andZEUScuts}. 

\begin{table}[ht]
\centering
\begin{tabular}{| c | c |}
\hline
 H1 cuts& ZEUS cuts \\\hline
$4 < Q^{2} < 110 \, \rm GeV^{2} $ & $Q^{2} > 25 \, \rm GeV^{2} \,$ \\ 
$x_\Pom < 0.1 $ & $x_{\Pom} < 0.01$ \\ 
$0.05 < y < 0.7 $ & $0.1 < y < 0.64$ \\ 
$ -1 < \eta_{1,2} < 2.5 $ & $\eta_{1,2} < 2$ \\ 
$p_{\perp 1} > 5 \, \rm GeV   $ & $p_{\perp 1,2}>2 \, \rm GeV$ \\ 
$p_{\perp 2} > 4 \, \rm GeV \,$ & $M_{jj} > 5 \, \rm GeV \,$ \\ 
$\left|t\right| < 1 \, \rm GeV^{2}$ &\, $90 < W_{\gamma p} < 250 \, \rm GeV \,$ \\\hline 
\end{tabular}
\caption{\label{tab:H1ZEUS}Cuts used by the H1 and ZEUS collaborations. }
\label{tab:H1andZEUScuts}
\end{table}

    In Table~\ref{tab:cross_sections} we present our phase space integrated cross sections for five different models of GTMDs. Quite different results are obtained for the different models. Results for light $q \bar{q}$ and $c \bar{c}$ dijets production are presented separately. For each model, the $c \bar{c}$ contribution is between 50 and 70$\%$ of the cross section of the light quark dijets cross section. We placed in this table also experimental cross sections as measured by the H1 and ZEUS collaborations. 
    All GTMDs except the MV-BS model underpredict experimental data,
    while the latter model is in the ballpark of H1 data but dramatically overpredicts the ZEUS results. 
    Not taking into account cuts on jet momenta $p_{\perp 1,2}$ results in a several-fold increase in the total cross-sections.  
    Here the $x_{\Pom}$-independent GTMD \cite{Boer:2021upt} was used.\\

\begin{table}[h]
\centering
\begin{tabular}{| l | c | c | c | c | c | c |}
\hline
GTMD  & H1, & H1, & H1, & ZEUS, & ZEUS, & ZEUS,\\
$\,$approach $\,$ & $\,$ light $q\bar{q}$ $\,$& $\,$ $ep \to  c\bar{c}$ $\,$ &no $p_{\perp 1,2}$ cuts & $\,$ light $q\bar{q}$ $\,$& $\,$ $ep \to  c\bar{c}$ $\,$ &no $p_{\perp 1,2}$ cuts \\
 & $\sigma$ $(pb)$ & $\sigma$ $(pb)$ & $\sigma$ $(pb)$ & $\sigma$ $(pb)$ & $\sigma$ $(pb)$ & $\sigma$ $(pb)$  \\\hline
GBW & 26.35 & 19.91 & 10900.86 & 13.57  &6.67 & 337.11\\ 
MPM & 147.94 & 108.26 & 10151.00 &  43.61 & 20.47 & 313.17\\ 
MV-BS & 404.06 & 269.75 & 10999.73 & 1346.11  & 624.55  & 3117.95 \\
KT & 21.29 & 15.20 & 5957.65& 12.57 & 5.67 & 52.60\\
MV-IR & 243.20 & 155.21 & 11784.75 & 37.83  & 17.62  & 91.18  \\ \hline
DATA & \multicolumn{2}{|c|}{254} & - & \multicolumn{2}{|c|}{72} & - \\ \hline
\end{tabular}
\caption{\label{tab:widgets}Total cross section for H1 and ZEUS conditions and different approaches.}
\label{tab:cross_sections}
\end{table}

\begin{figure}
  \centering
  \includegraphics[width=0.47\textwidth]{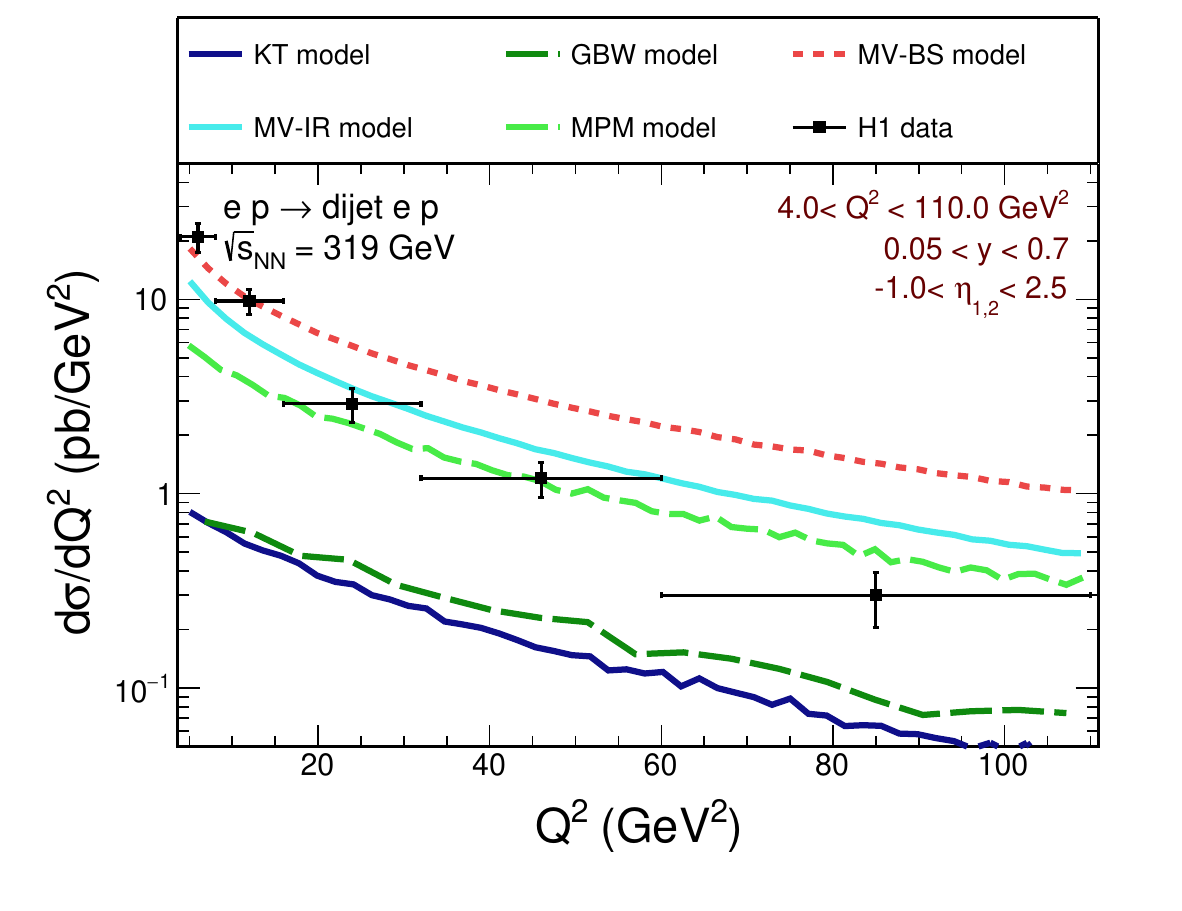}
  \includegraphics[width=0.47\textwidth]{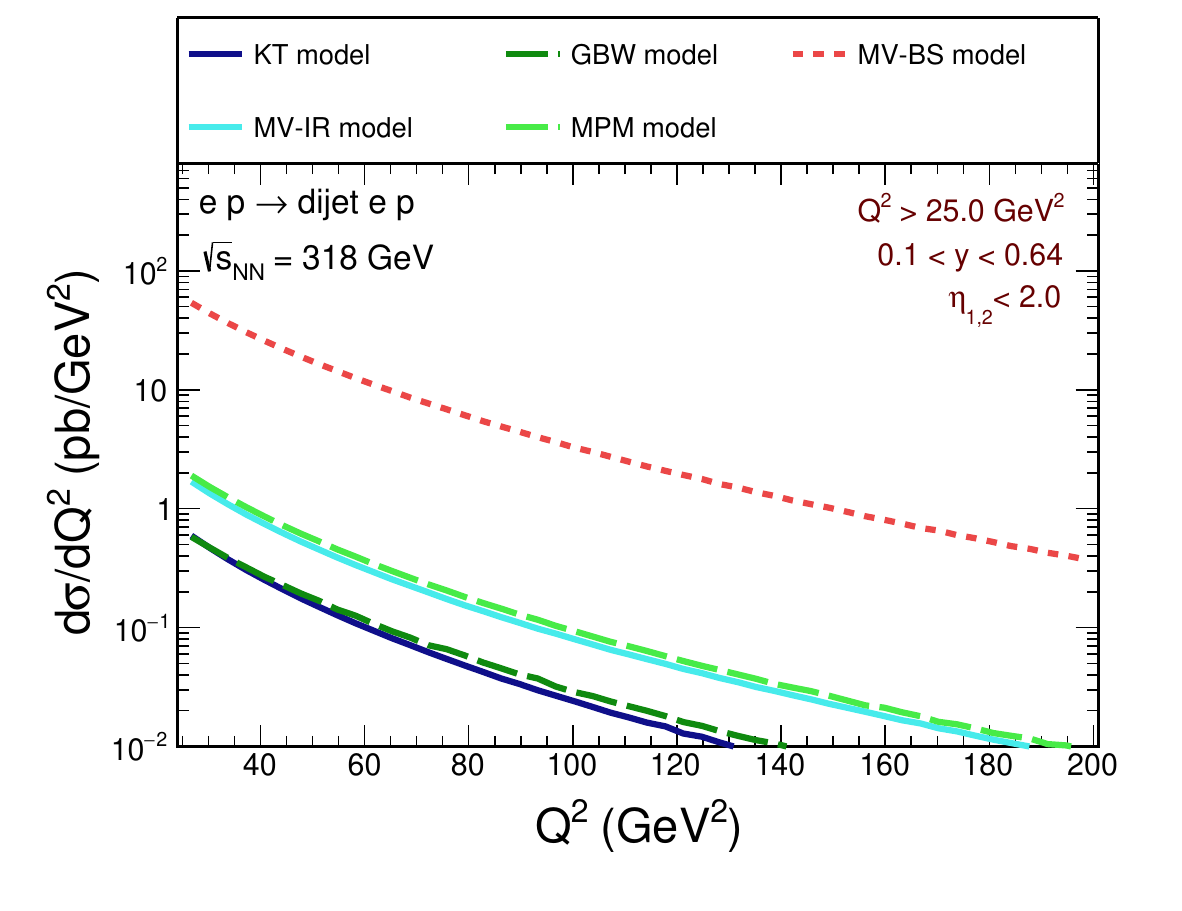}
  \caption{$Q^{2}$ dependence of the cross section for H1 (left) and ZEUS (right) kinematics for different GTMDs.}
\label{dsig_dQ2}  
\end{figure}

\begin{figure}
  \centering
  \includegraphics[width=0.47\textwidth]{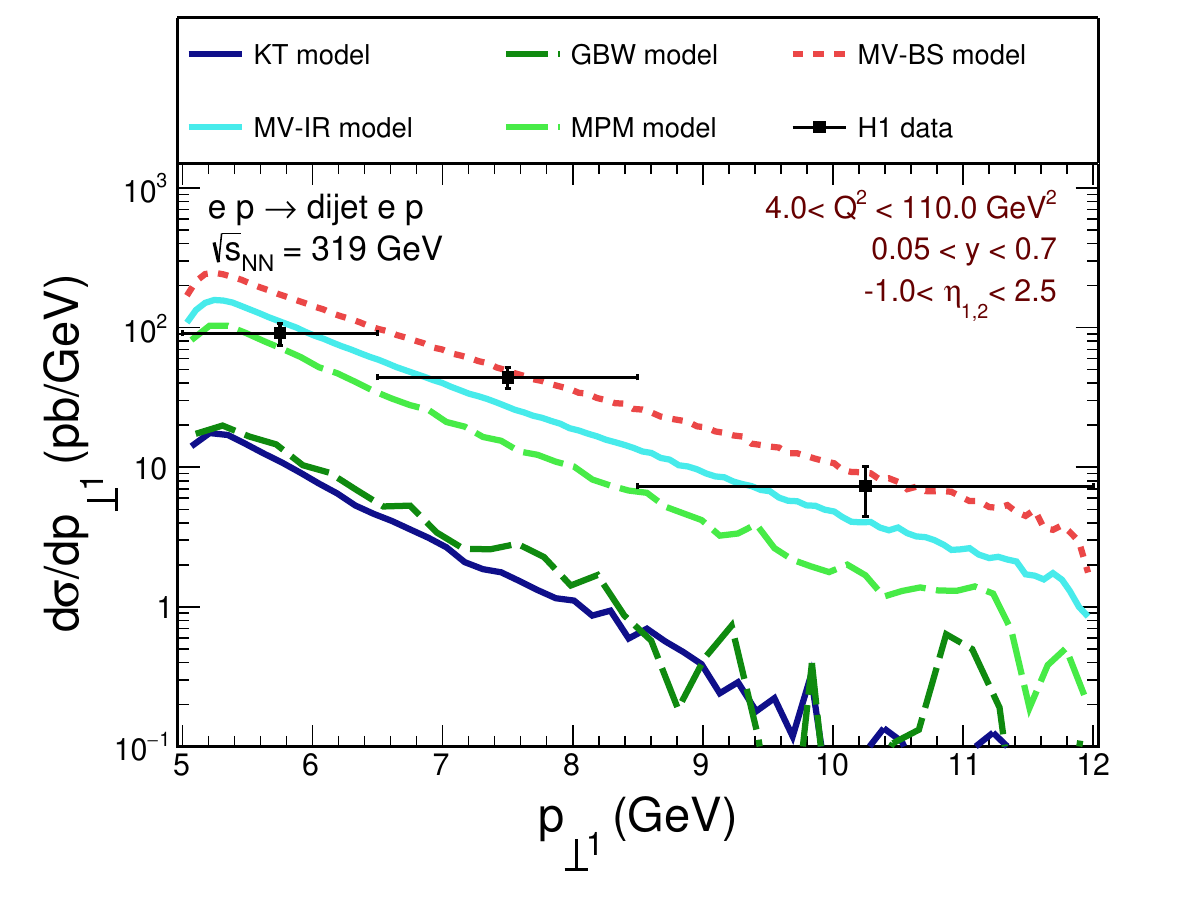}
  \includegraphics[width=0.47\textwidth]{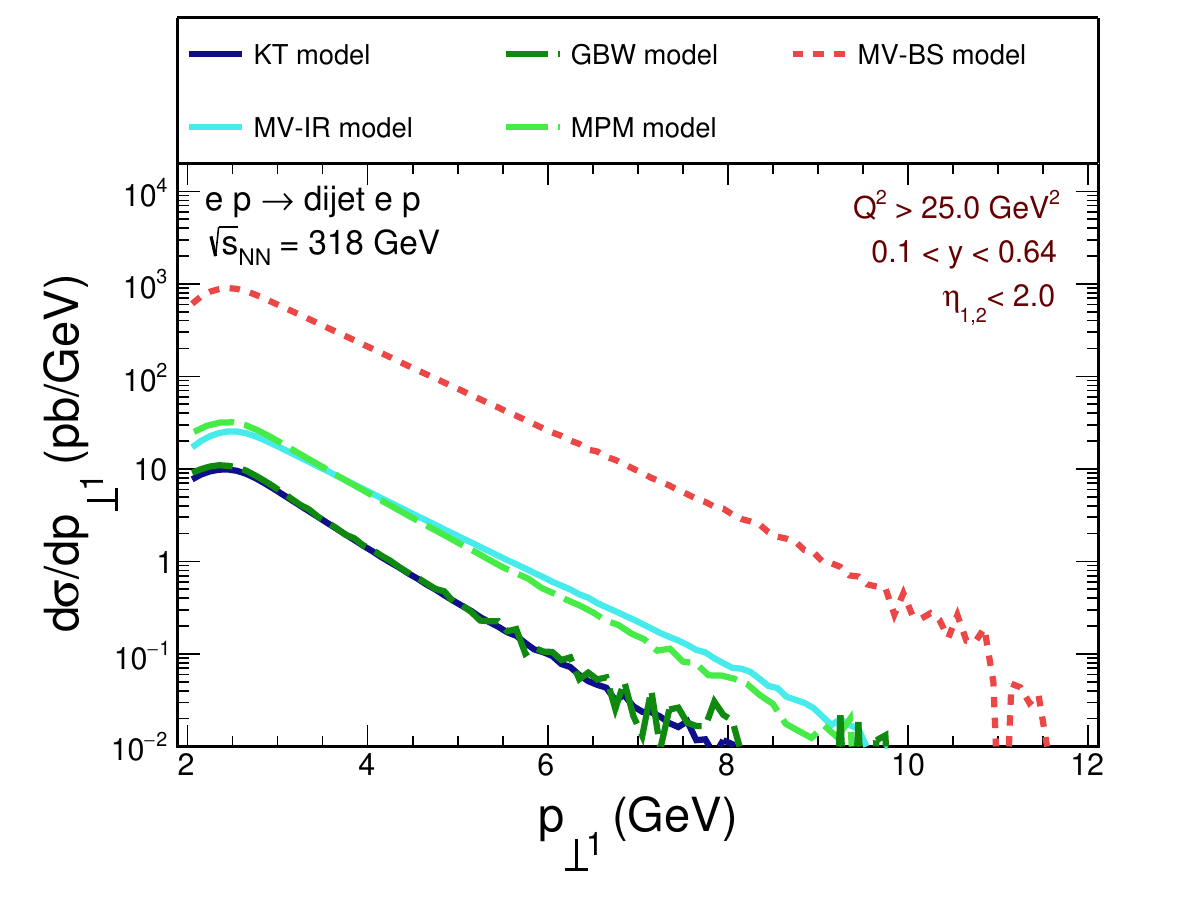}
  \caption{Jet transverse momentum dependence of the cross section for H1 (left) and ZEUS (right) kinematics for different GTMDs.}
\label{dsig_dpt1}
\end{figure}

\begin{figure}
  \centering
  \includegraphics[width=0.47\textwidth]{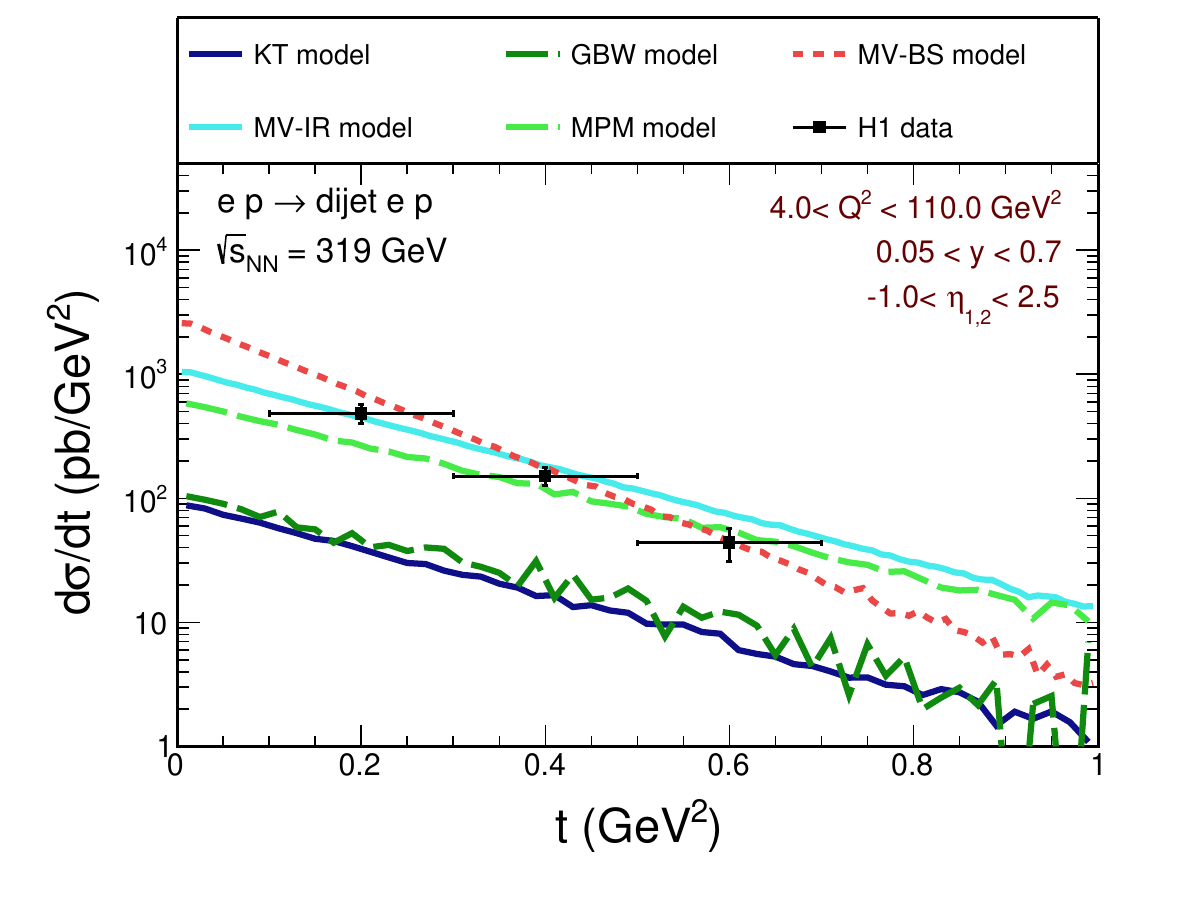}
  \includegraphics[width=0.47\textwidth]{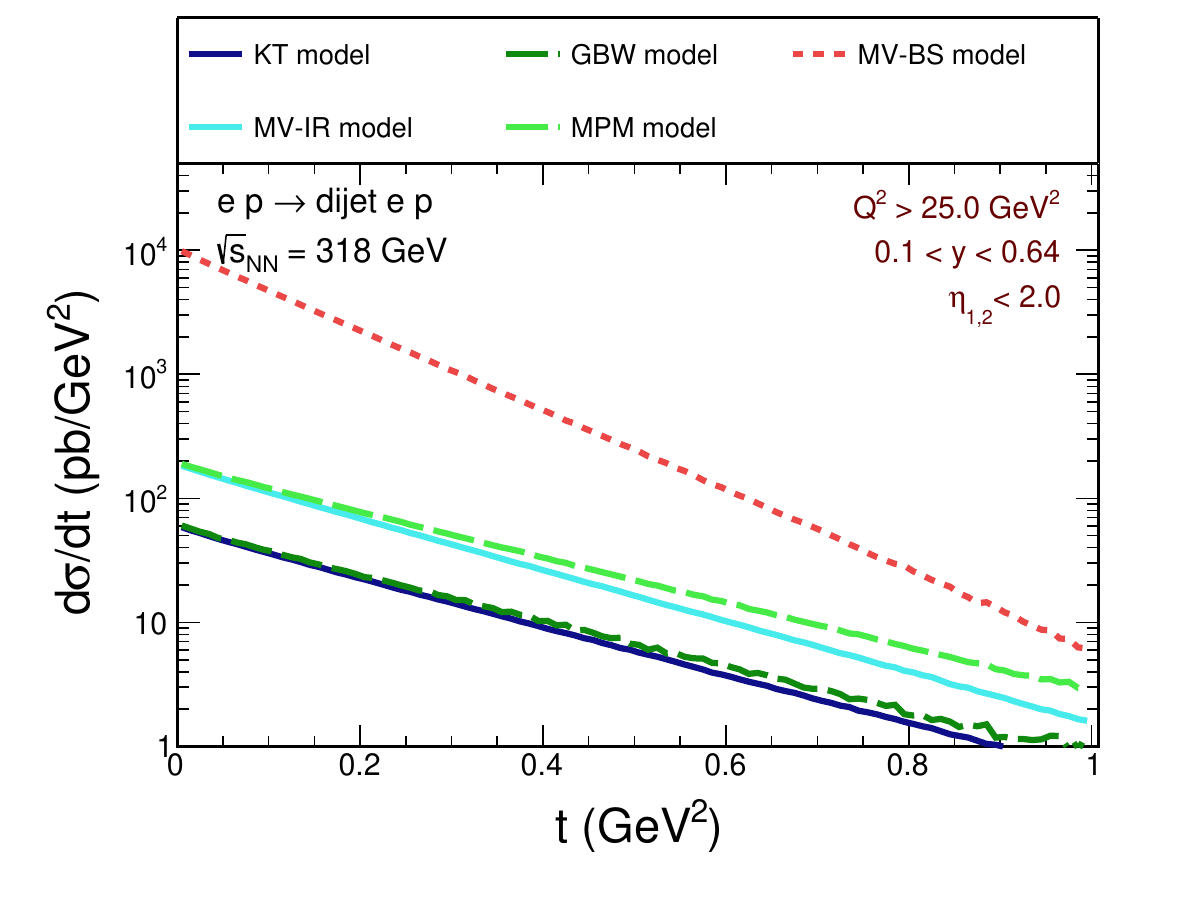}
  \caption{Mandelstam $t$ dependence of of the cross section for H1 (left) and ZEUS (right) kinematics for different GTMDs.}
\label{dsig_dt}
\end{figure}

\begin{figure}
  \centering
  \includegraphics[width=0.47\textwidth]{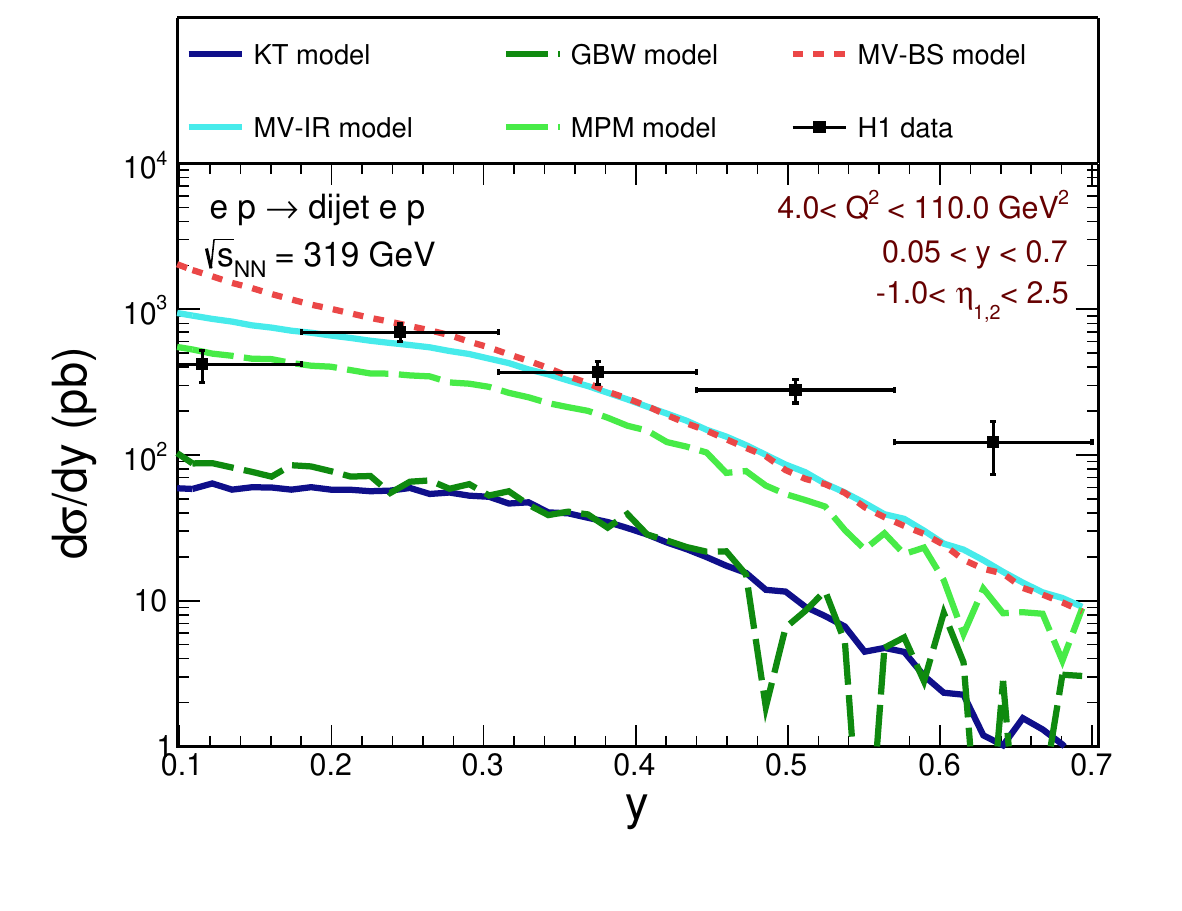}
  \includegraphics[width=0.47\textwidth]{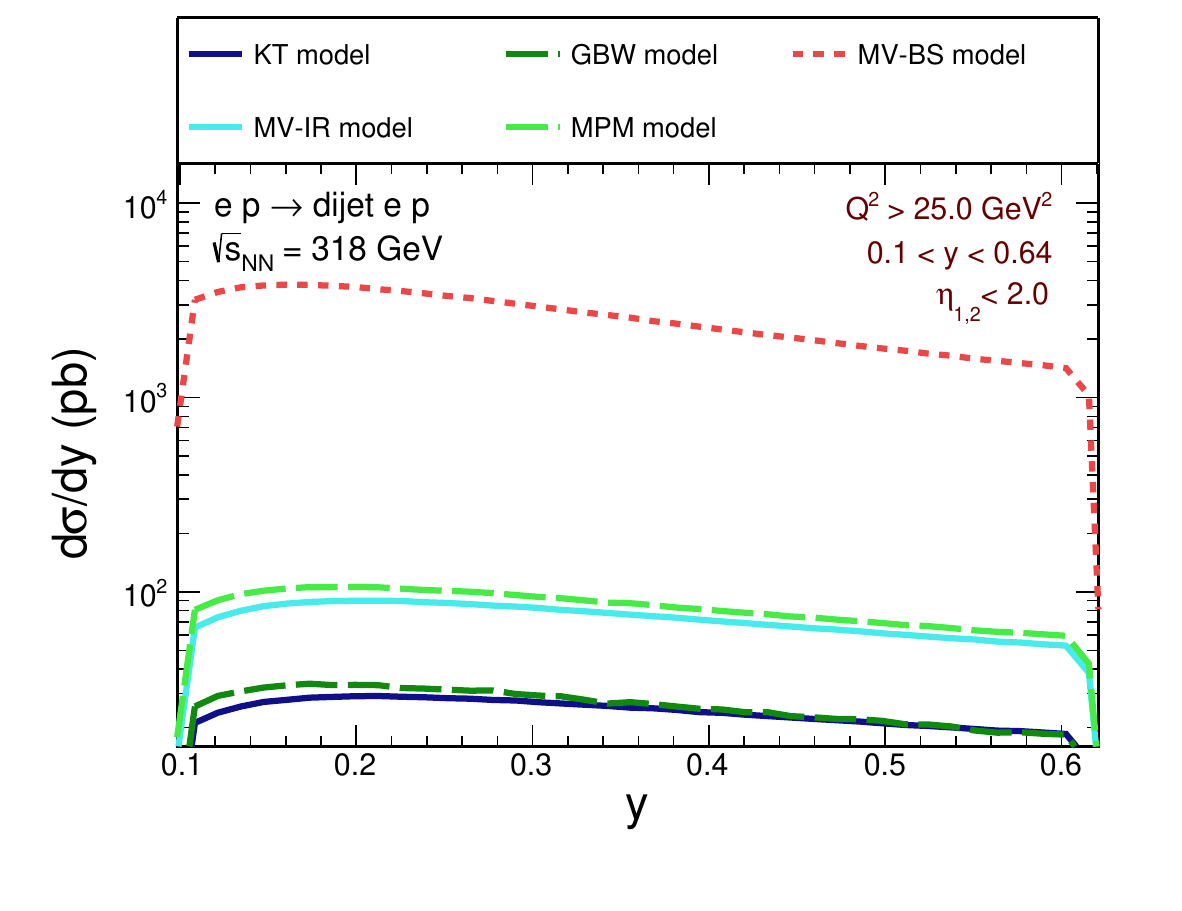}
  \caption{Inelasticity dependence of the cross section for H1 (left) and ZEUS (right) kinematics for different GTMDs.}
\label{dsig_dyin}
\end{figure}

\begin{figure}
  \centering
  \includegraphics[width=0.47\textwidth]{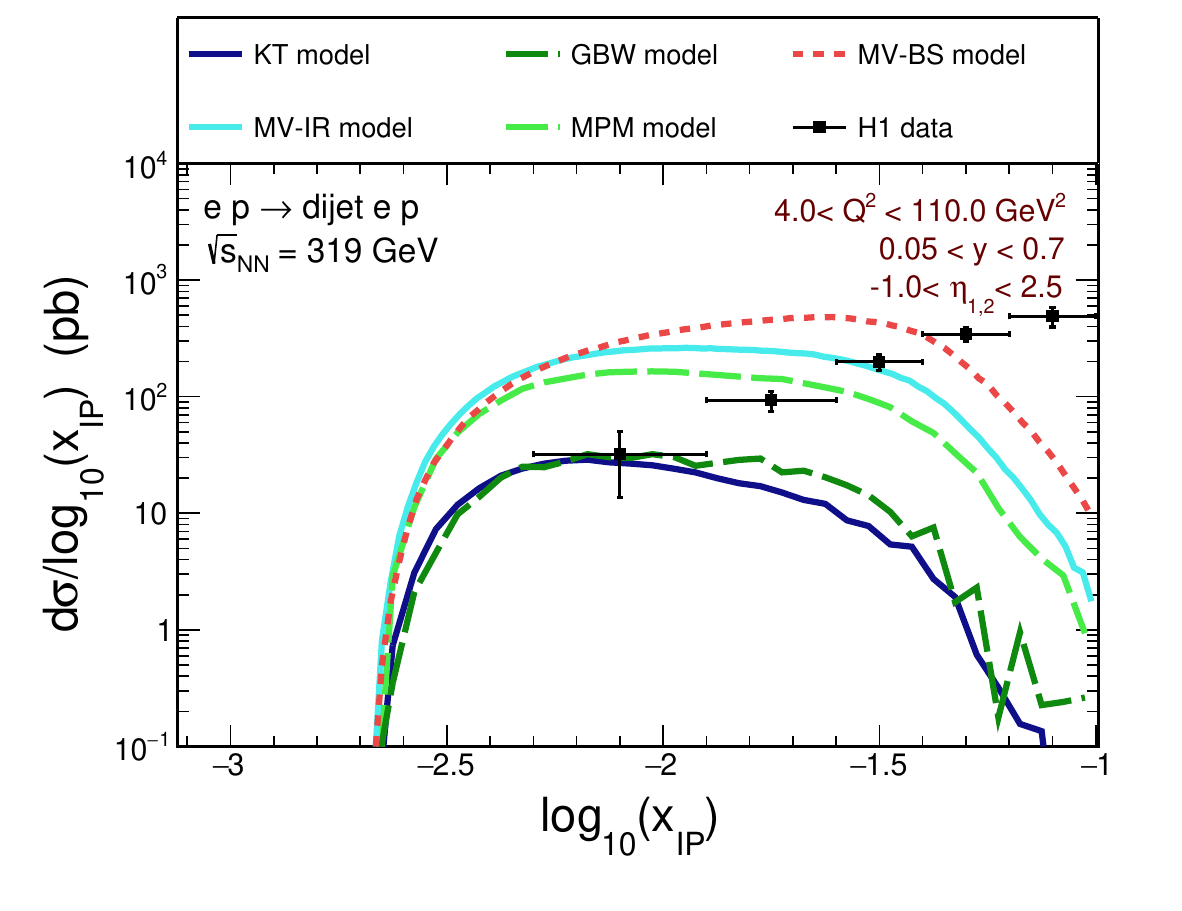}
  \includegraphics[width=0.47\textwidth]{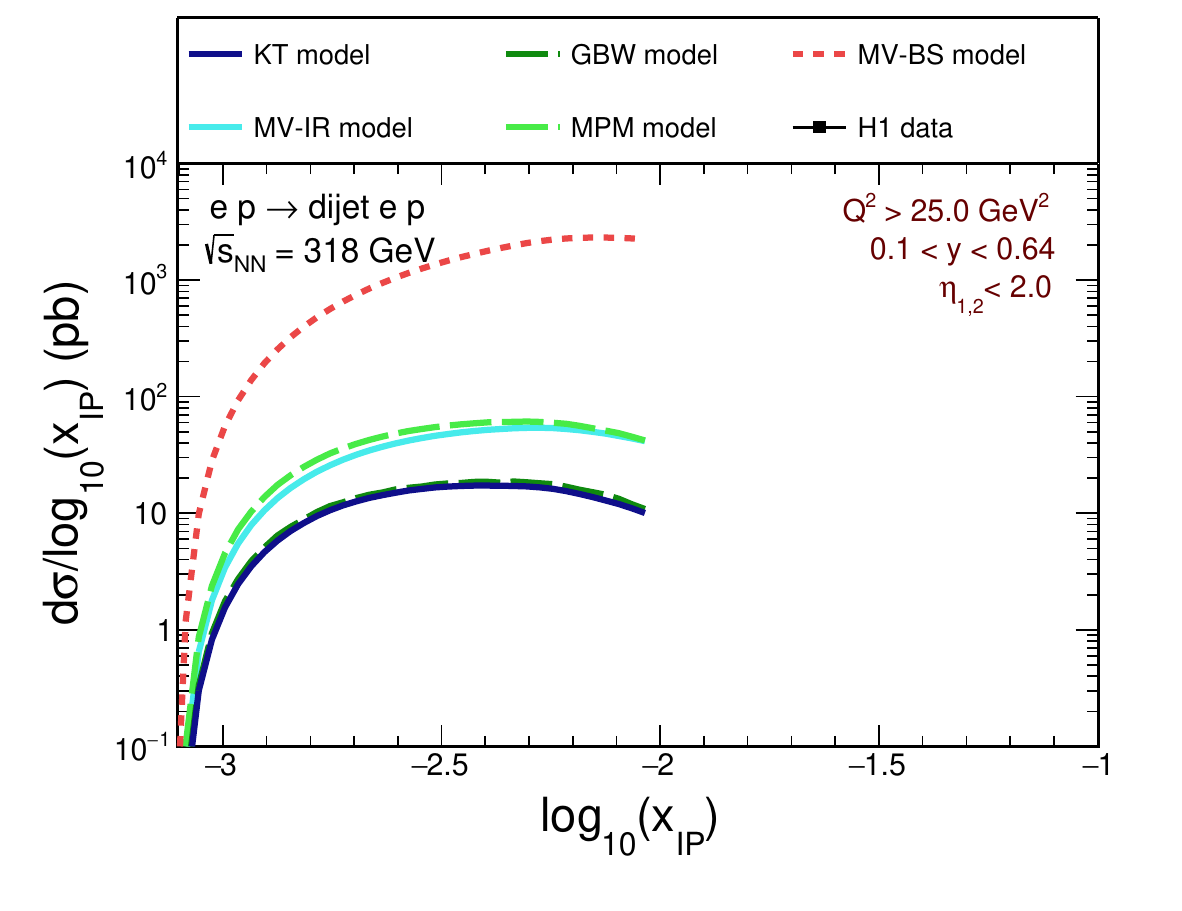}
  \caption{$x_{\Pom}$ distribution dependence of the cross section for H1 (left) and ZEUS (right) kinematics for different GTMDs.}
\label{dsig_dlog10xp}
\end{figure}

\begin{figure}
  \centering
  \includegraphics[width=0.47\textwidth]{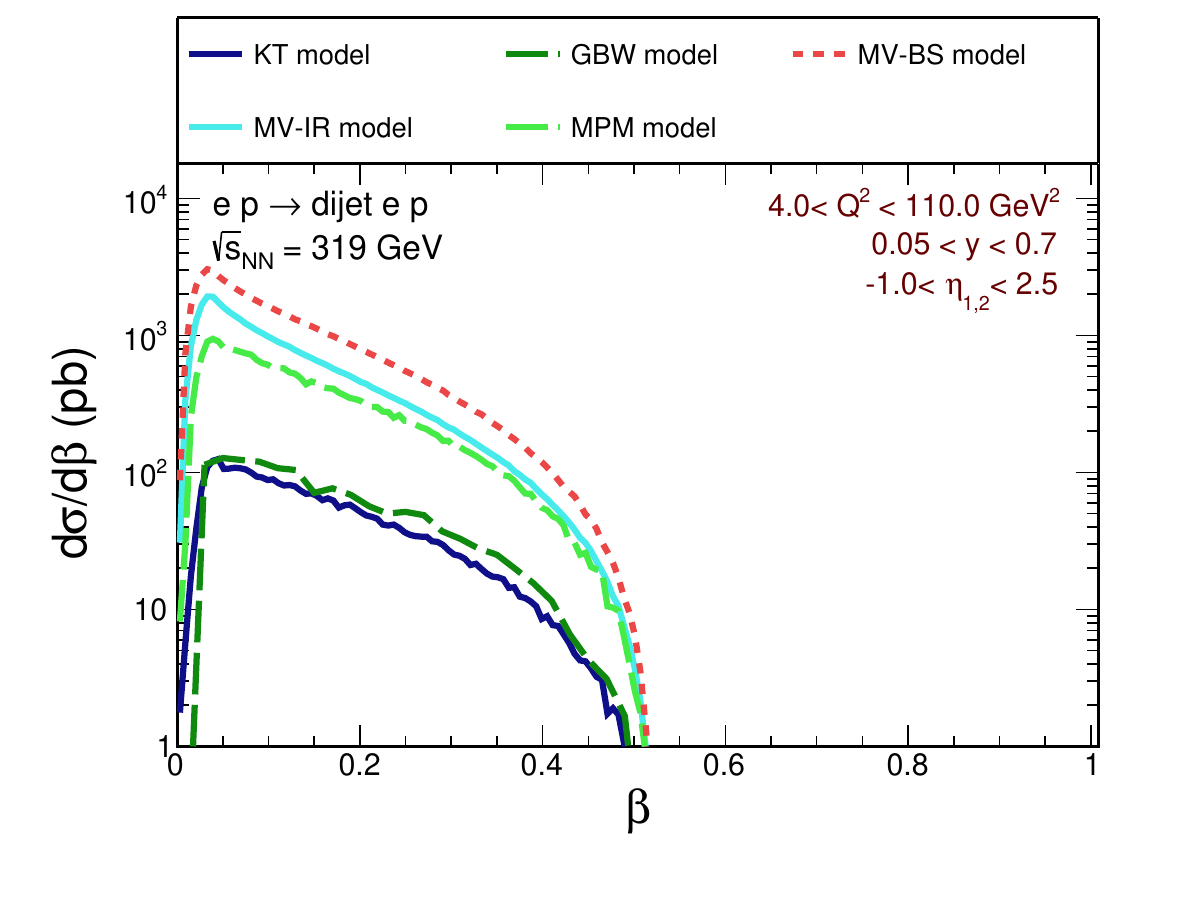}
  \includegraphics[width=0.47\textwidth]{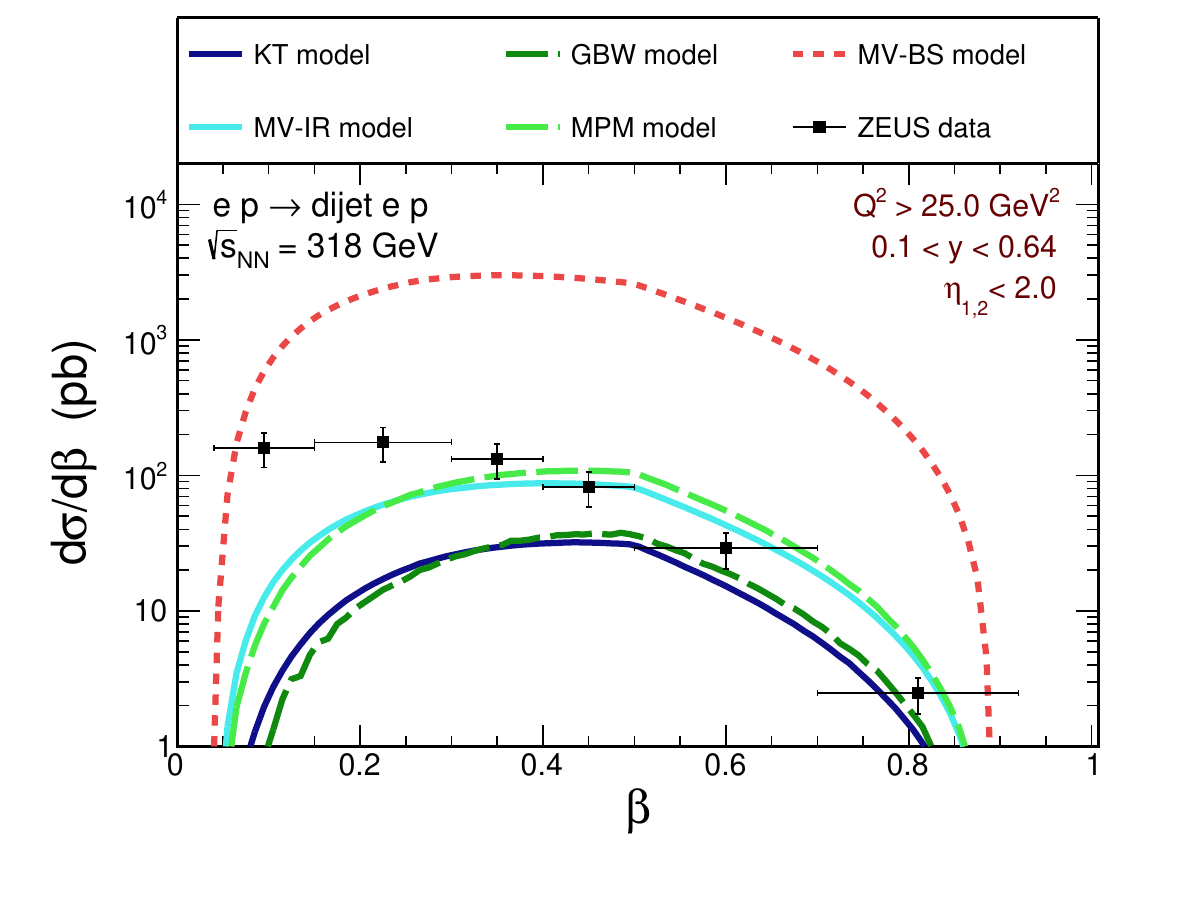}
\caption{$\beta$ dependence of the cross section for H1 (left) and ZEUS (right) kinematics for different GTMDs.}
\label{dsig_dbeta}
\end{figure}

We now turn to the discussion of distributions in various kinematical variables.
We show distributions in the $Q^2$, jet transverse momentum, Mandelstam $t$, inelasticity $y$, pomeron $x$ and standard diffractive variable $\beta$ both for H1 and ZEUS kinematics along with the corresponding experimental data.
In Fig.~\ref{dsig_dQ2}, we show the cross section differential in $Q^2$. At large $Q^2 > 20 \, \rm{GeV}^2$, for all GTMDs, the shape of the cross section reasonably well agrees with the H1 data. However there are large differences in the normalization of the cross section, as is evident already from Table \ref{tab:cross_sections}. The largest contribution is obtained for the MV-BS model. For the kinematics of the ZEUS data, the MV-BS result differs by almost two orders of magnitude from other GTMDs.
A similar good agreement of the shapes of distributions is observed for the jet transverse momentum, see Fig.\ref{dsig_dpt1}.
However in the $t$ distribution shown in Fig. \ref{dsig_dt} the slope of MV-BS model is slightly different from the other GTMDs.The Boer-Setyadi GTMD gives results that are close to the H1 data. So at first look the distributions seems to be supported by the data.
The distribution in inelasticity (or electron energy loss) $y$ is shown in Fig.\ref{dsig_dyin}. The shape is reasonably well reproduced for all distributions, although the falloff towards large $y$ is a bit too strong.

The situation changes for the distribution in $\log_{10} x_\Pom$, shown in Fig.~\ref{dsig_dlog10xp}. 
notice, that $\Delta Y \equiv \ln(1/x_\Pom)$ is essentially the rapidity gap between proton and diffractive system. Here the MV-BS distribution is evidently disfavoured by the H1 data, similarly to the MV-IR and MPM models whereas the GBW and KT GTMDs give cross section which is not in contradiction with the data. It is worth noting that in the H1 kinematics  we have rather large $x_{\Pom}>0.01$.
Here one does not expect the Pomeron exchange to dominate, but rather the contribution of $q \bar q$ exchanges described by the secondary Reggeons will take over. See for example the case of $pp$
scattering in Ref. \cite{Luszczak:2014cxa}, where secondary Reggeons give a sizeable contribution. Therefore, it should not be expected that all experimental data will be correctly reproduced by the dipole approach which is appropriately used for, say, small $x_{\Pom} < 0.01$.
In fact is the ZEUS data for which one can expect the Pomeron or gluon GTMD contribution to be appropriate.

The ZEUS collaboration has measured the distribution in $\beta$, which we show for the kinematic cuts of both H1 and ZEUS in Fig.\ref{dsig_dbeta}. Here we see, that the
MV-BS GTMD dramatically overpredict the ZEUS data while MPM and MV-IR do not exceed the values of experimental data within their measurement uncertainties. 
It appears that the Boer-Setyadi GTMD distribution is artificially large in order to describe the H1 data, where gluons arguably do not dominate. 
The distributions in $x_{\Pom}$ and $\beta$ show that the agreement of  the MV-BS model above is rather artificial.
We caution the reader, that even for the ZEUS data it would likely be inappropriate to fit the GTMD to data using the present approach. There are many indications \cite{ZEUS:2016}, that the 
$q \bar q g$ three-parton contribution is important in this context (see e.g. Ref. \cite{Bartels:1999tn}). It seems that the KT and GBW models are the most realistic ones.

\begin{figure}
  \centering 
  \includegraphics[width=0.47\textwidth]{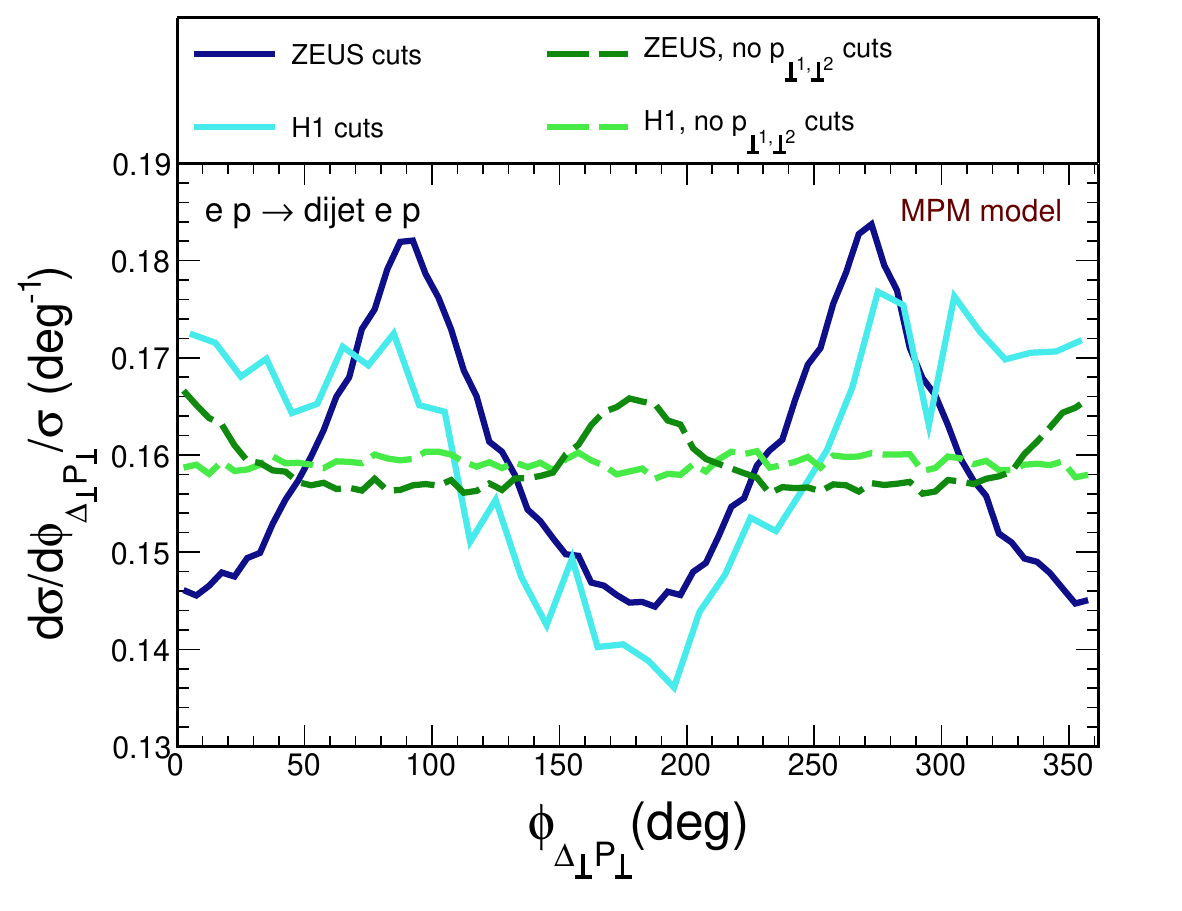} 
  \includegraphics[width=0.47\textwidth]{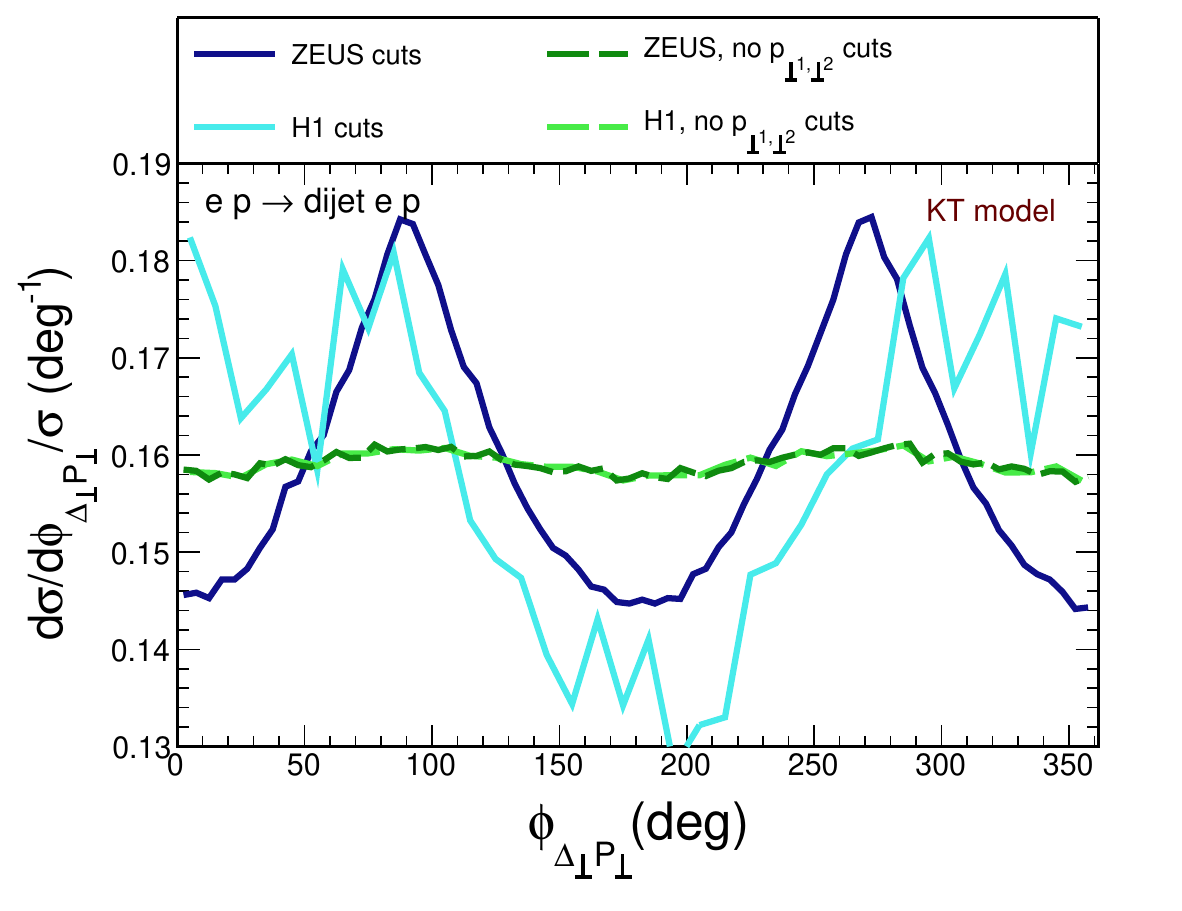} 
  \includegraphics[width=0.47\textwidth]{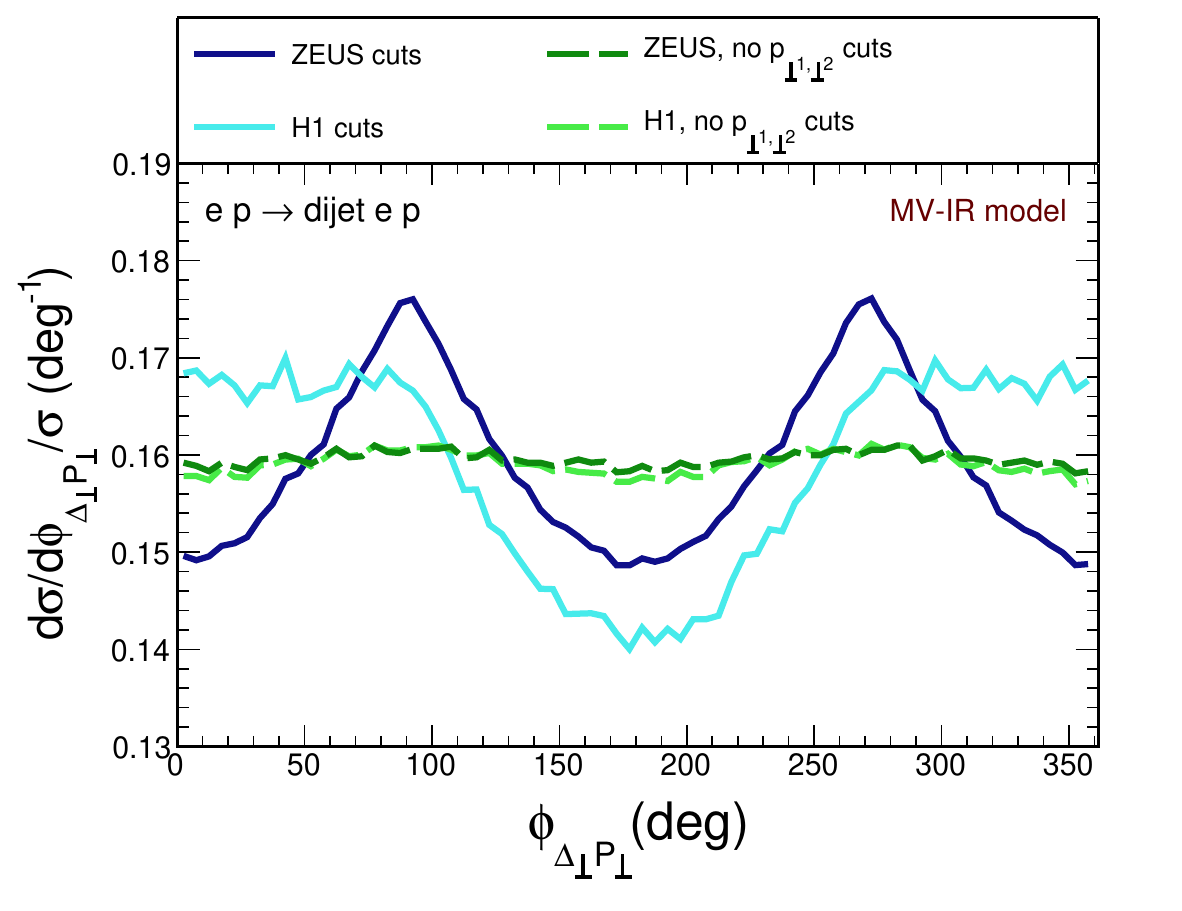}
  \includegraphics[width=0.47\textwidth]{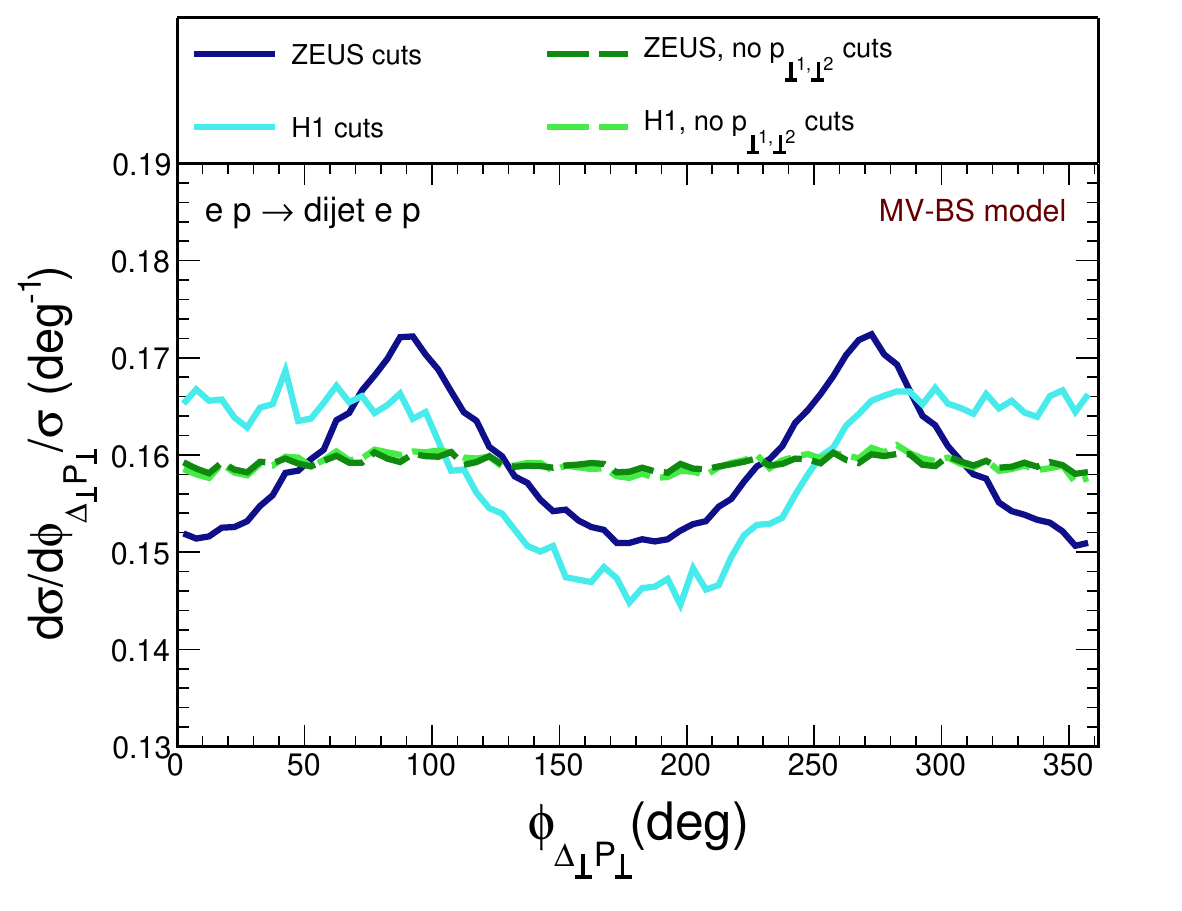}
  \caption{Distributions in the azimuthal angle $\phi$ between $\vec P_\perp$ and $\vec \Delta_\perp$ normalised to the total cross section for the H1 and ZEUS kinematics for different GTMDs. }
 \label{dsig_dphi}
\end{figure}

For completeness in Fig.~\ref{dsig_dphi} we show distribution in azimuthal angle between the sum and the difference of the jet transverse momenta. Such studies were performed before for $c \bar c$ correlations in $pA$ collisions \cite{Linek:2023kga}. We predict interesting azimuthal correlations for dijets at HERA. The straight horizontal line corresponds to the case without cuts on jet transverse momenta. The figures show that the correlation is mainly due to the cuts. In principle, such azimuthal correlation can be also due to elliptical gluon distributions \cite{Linek:2023kga}, not taken into account in the present paper. The observed effect can be therefore easily misinterpreted as observation of an elliptical glue. Of course, the elliptical glue can generate further effect on top of azimuthal correlations generated by the cut effect.

\begin{figure}
  \centering
  \includegraphics[width=0.47\textwidth]{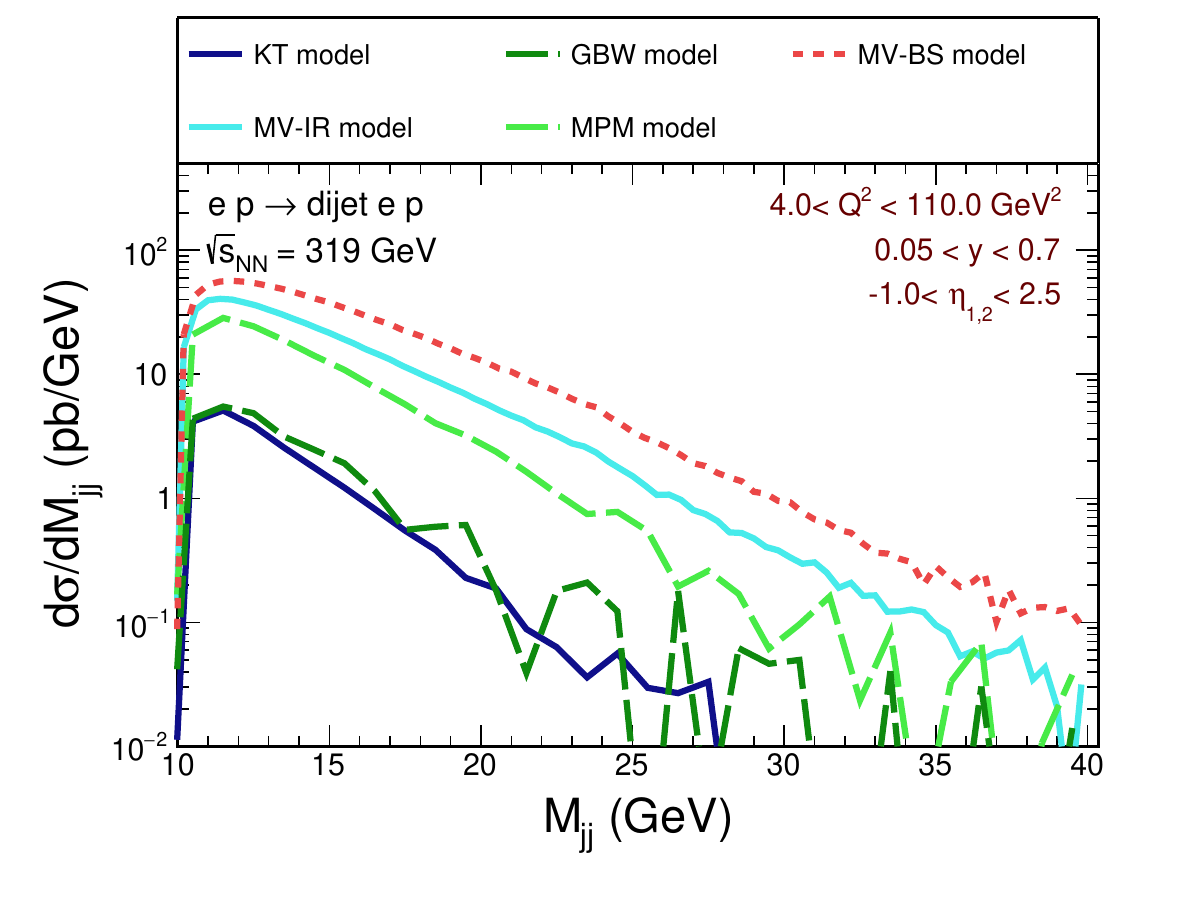}
  \includegraphics[width=0.47\textwidth]{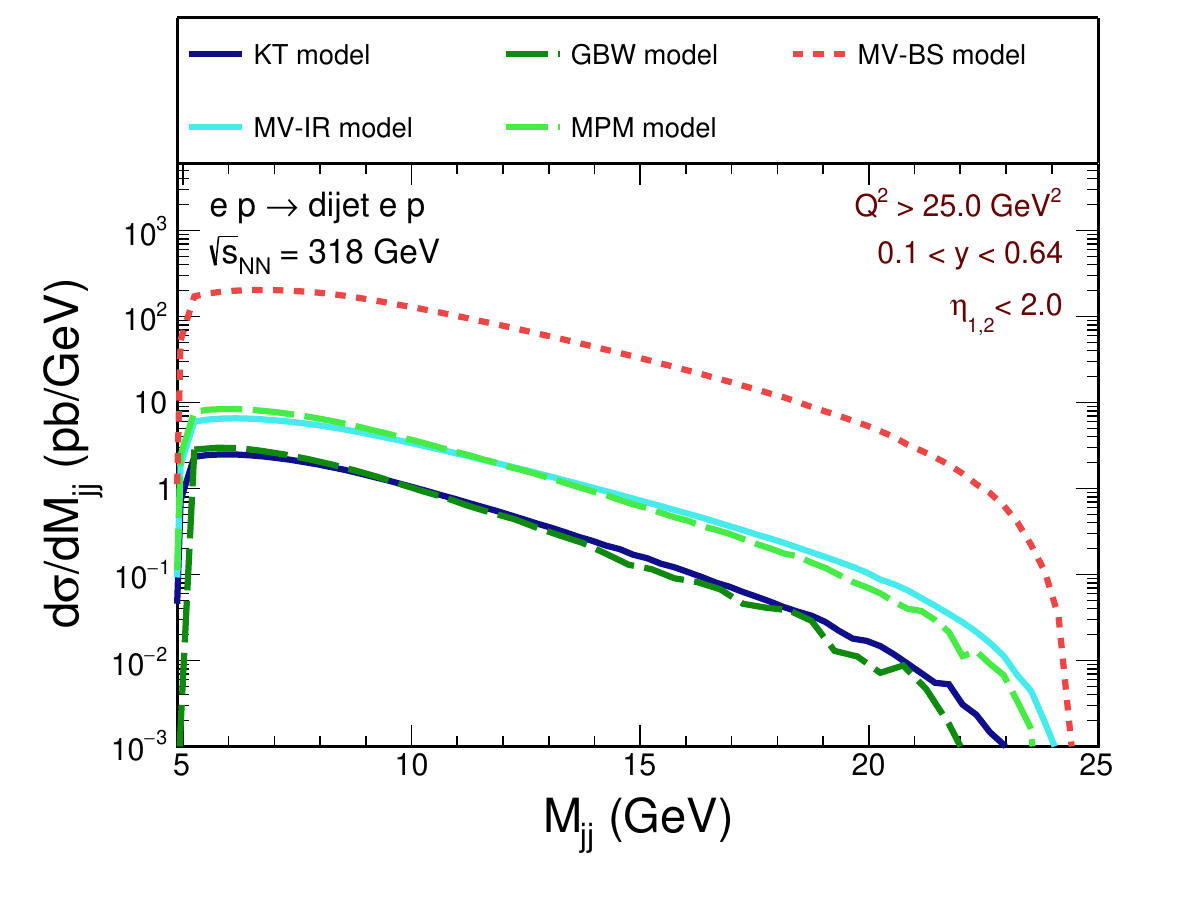}
  \caption{Invariant mass dependence of the cross section for H1 (left) and ZEUS (right) kinematics for different GTMDs.}
\label{dsig_dminv}
\end{figure}

\begin{figure}
  \centering
  \includegraphics[width=0.47\textwidth]{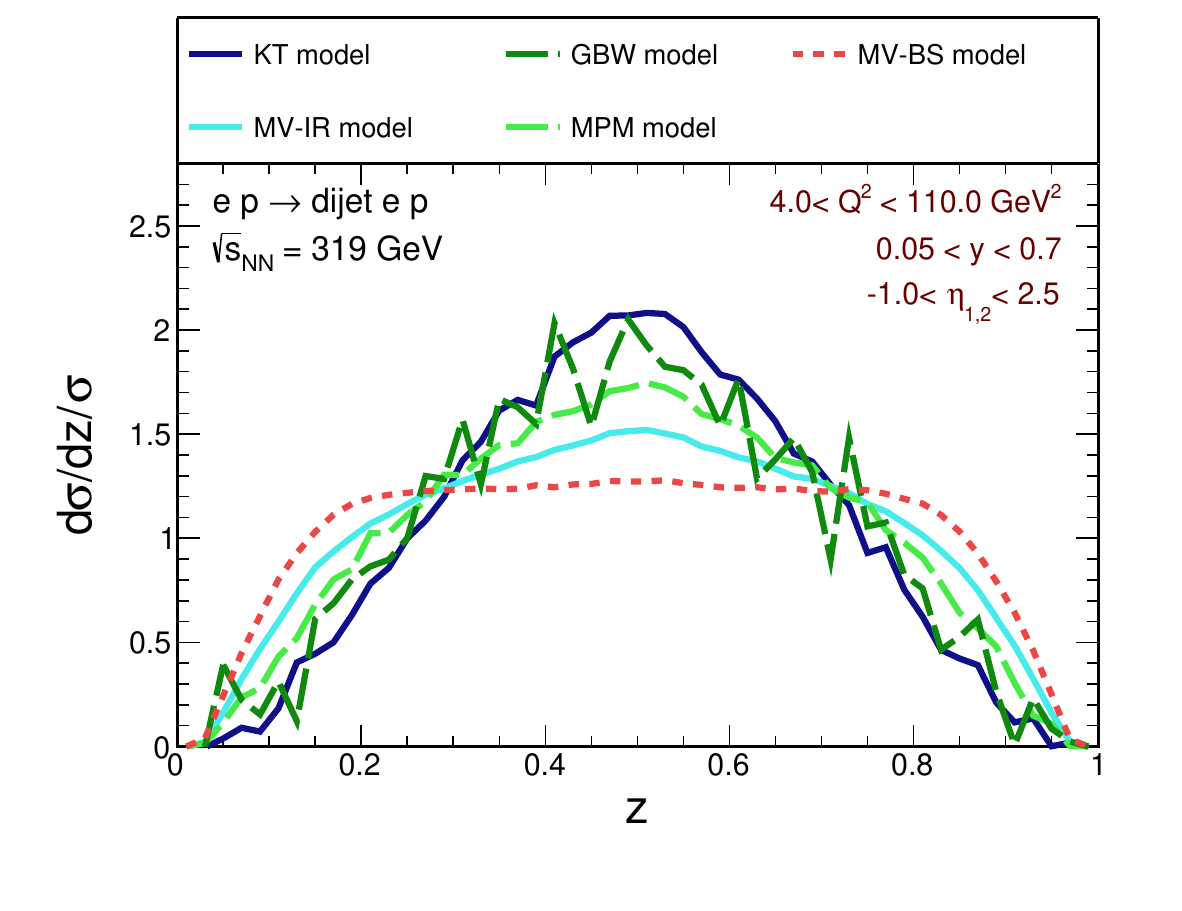}
  \includegraphics[width=0.47\textwidth]{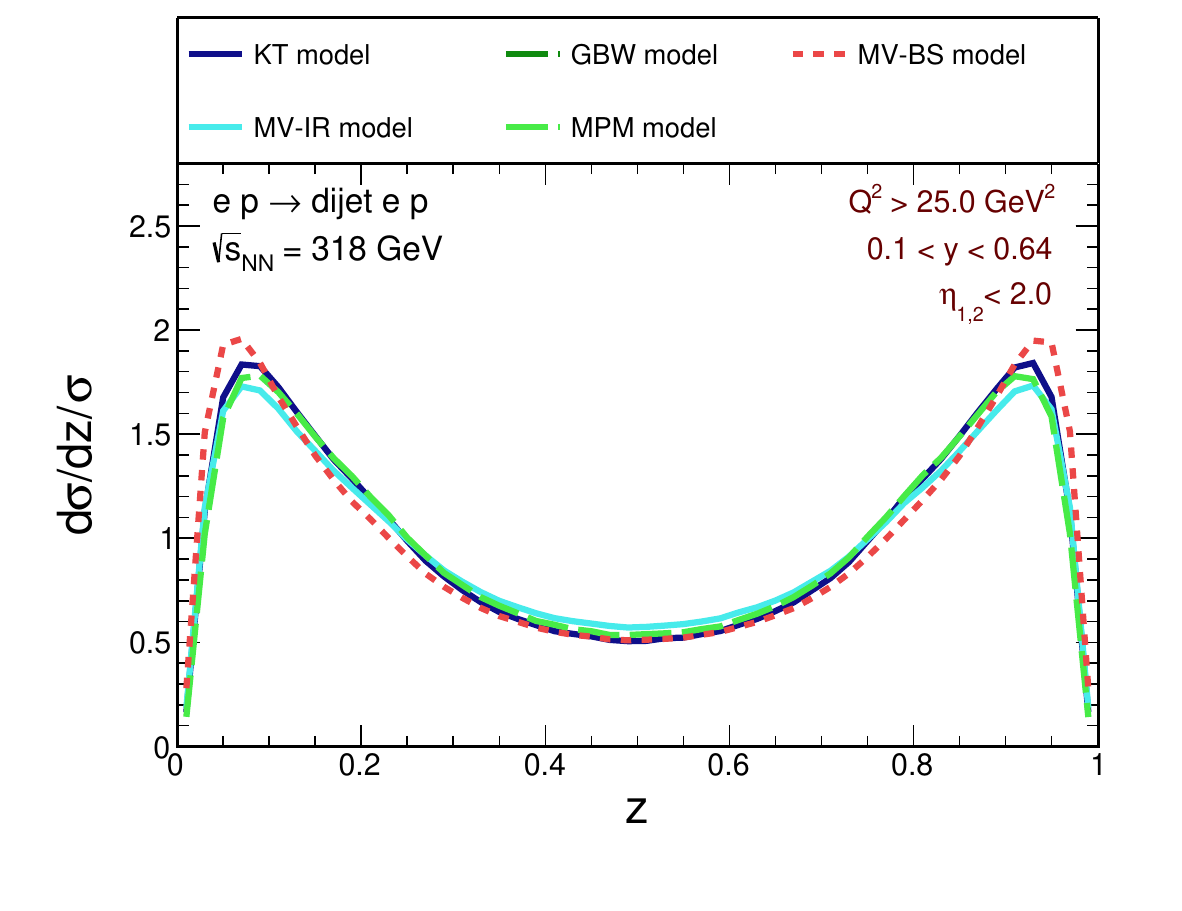}
  \caption{$z$ dependence of the cross section for H1 (left) and ZEUS (right) kinematics for different GTMDs. Please note the normalization.}
\label{dsig_dz}
\end{figure}

\begin{figure}
  \centering
  \includegraphics[width=0.47\textwidth]{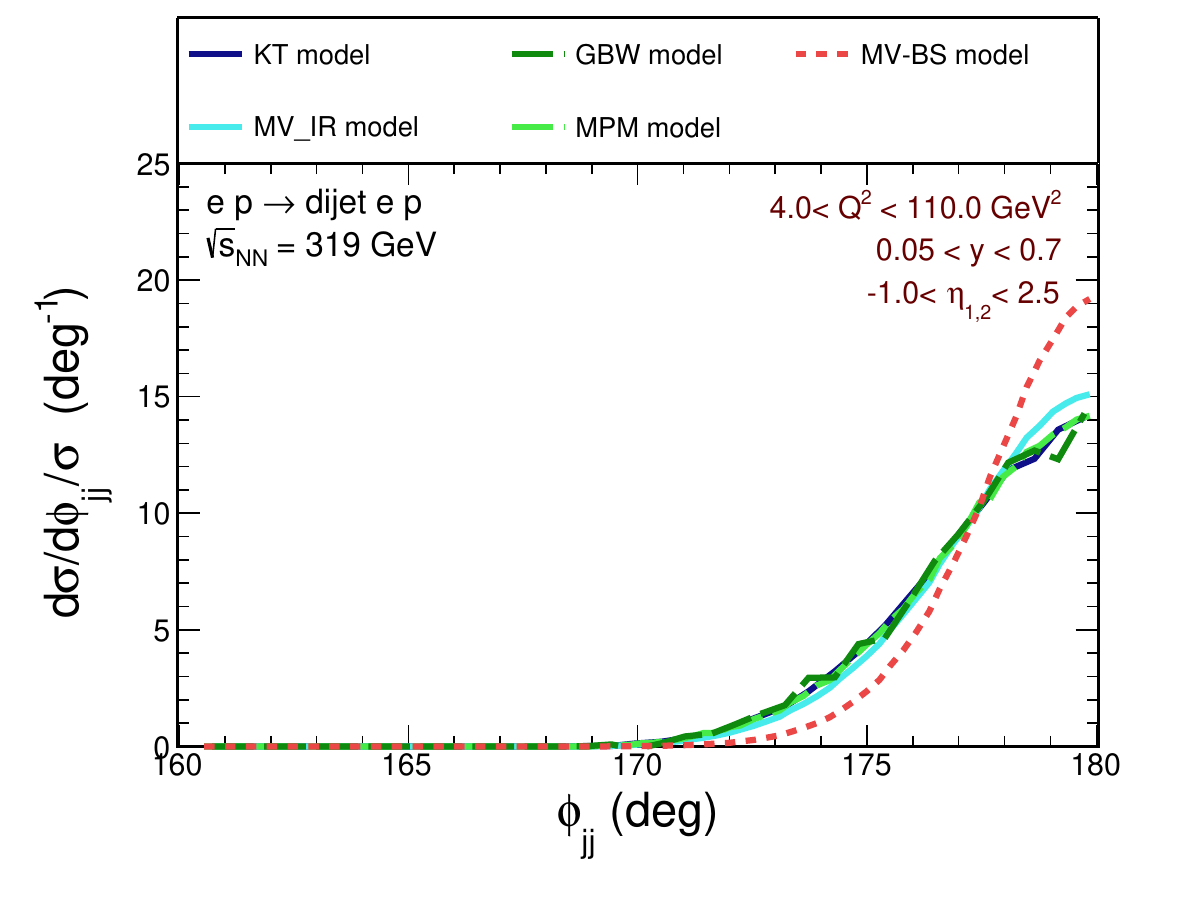}
  \includegraphics[width=0.47\textwidth]{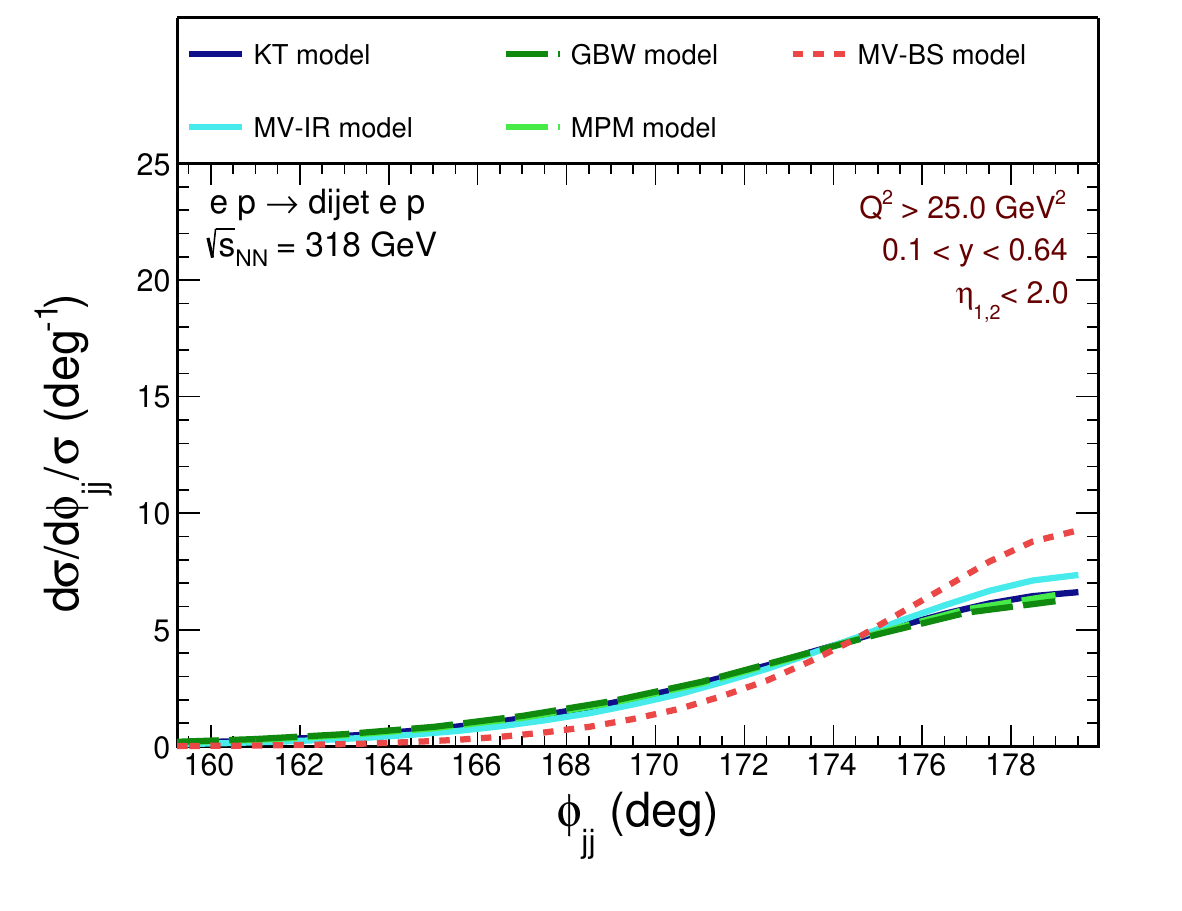}
\caption{Azimuthal angle between produced jets dependence of the cross section for H1 (left) and ZEUS (right) kinematics for different GTMDs. Please note the normalization.}
\label{dsig_dphijj}
\end{figure}

\begin{figure}
  \centering
  \includegraphics[width=0.47\textwidth]{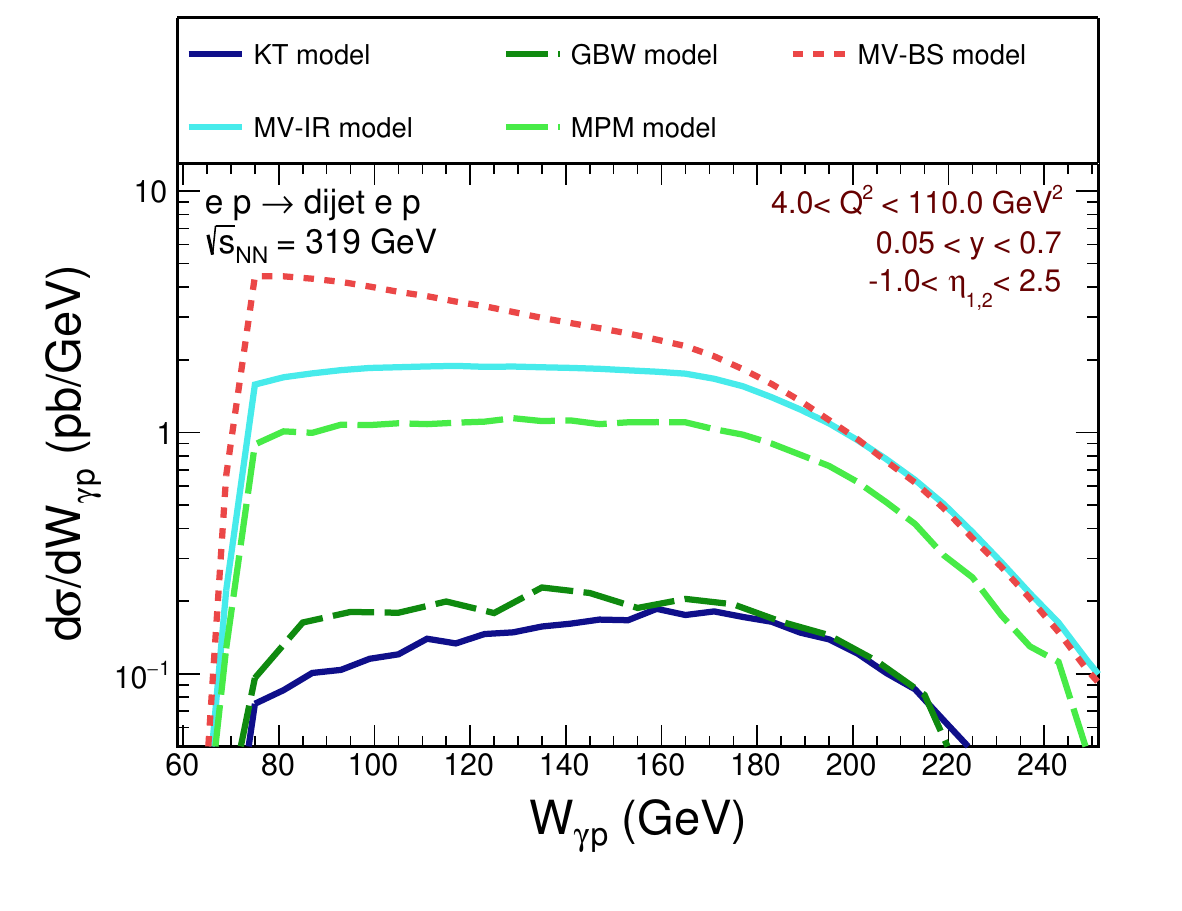}
  \includegraphics[width=0.47\textwidth]{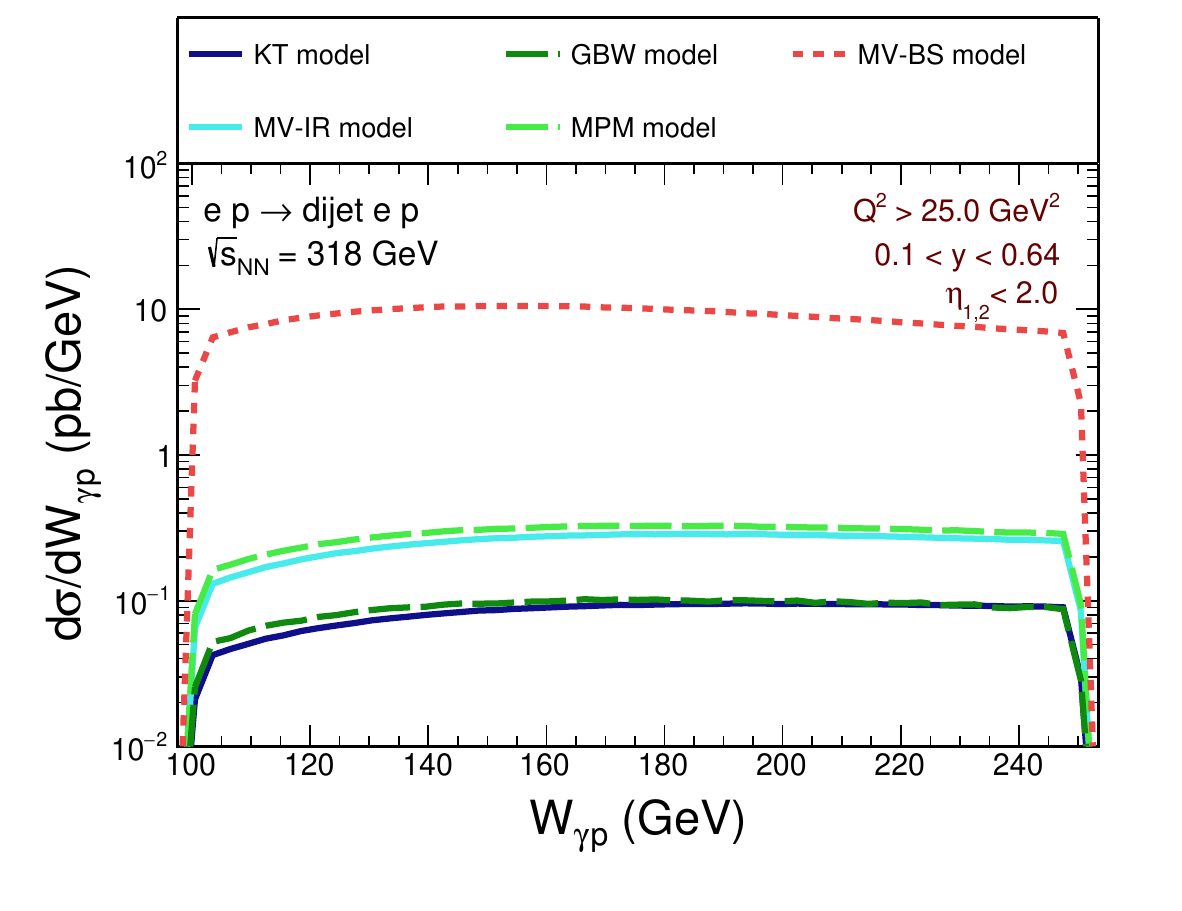}
  \caption{$W_{\gamma p}$ dependence of the cross section for H1 (left) and ZEUS (right) kinematics for different GTMDs.}
\label{dsig_dWgamp}
\end{figure}

In addition we show distributions in $M_{jj}$, $z$, $\phi_{jj}$ and $W_{\gamma p}$ in Fig.~\ref{dsig_dminv},~\ref{dsig_dz}, ~\ref{dsig_dphijj}, ~\ref{dsig_dWgamp}, respectively. 

The distributions in $M_{jj}$ are characterized by almost identical shapes for all GTMD models and both kinematic ranges, in contrast to the distribution in $z$, see Fig.~\ref{dsig_dz}. The maximum is visible for H1 cuts in this example at $z=0.5$ for all models except MV-BS, the distribution of which is almost flat in the range of $0.1-0.9$. The remaining models differ as to the value of the mentioned maximum. The situation is the opposite in the ZUES kinematics, where for $z= 0.5$ there is a minimum of the distributions of all models.\\
The high peak at $\phi_{jj}=180 \deg$ in the distribution in the azimuthal angle between produced jets for the H1 kinematics from Fig. ~\ref{dsig_dphijj} shows that the jets are almost back-to-back, while in the case of ZEUS cuts that effect is less pronounced. 

Taking into account the distributions discussed above, we conclude that the H1 \cite{H1:2012} and ZEUS \cite{ZEUS:2016} data contain mechanisms other than those shown in Fig.~\ref{fig:diagrams}, which will be the subject of further investigations.

\begin{figure}
  \centering 
  \includegraphics[width=0.47\textwidth]{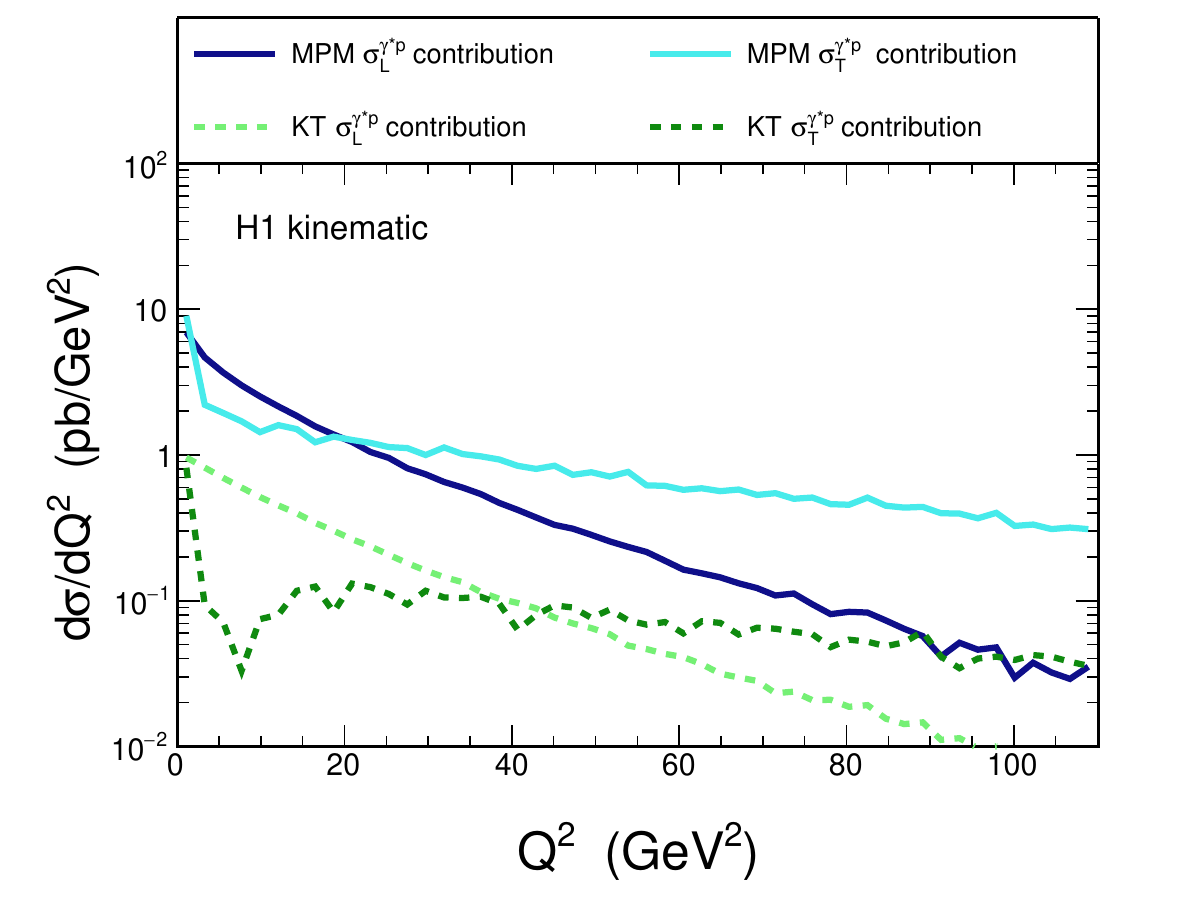}
  \includegraphics[width=0.47\textwidth]{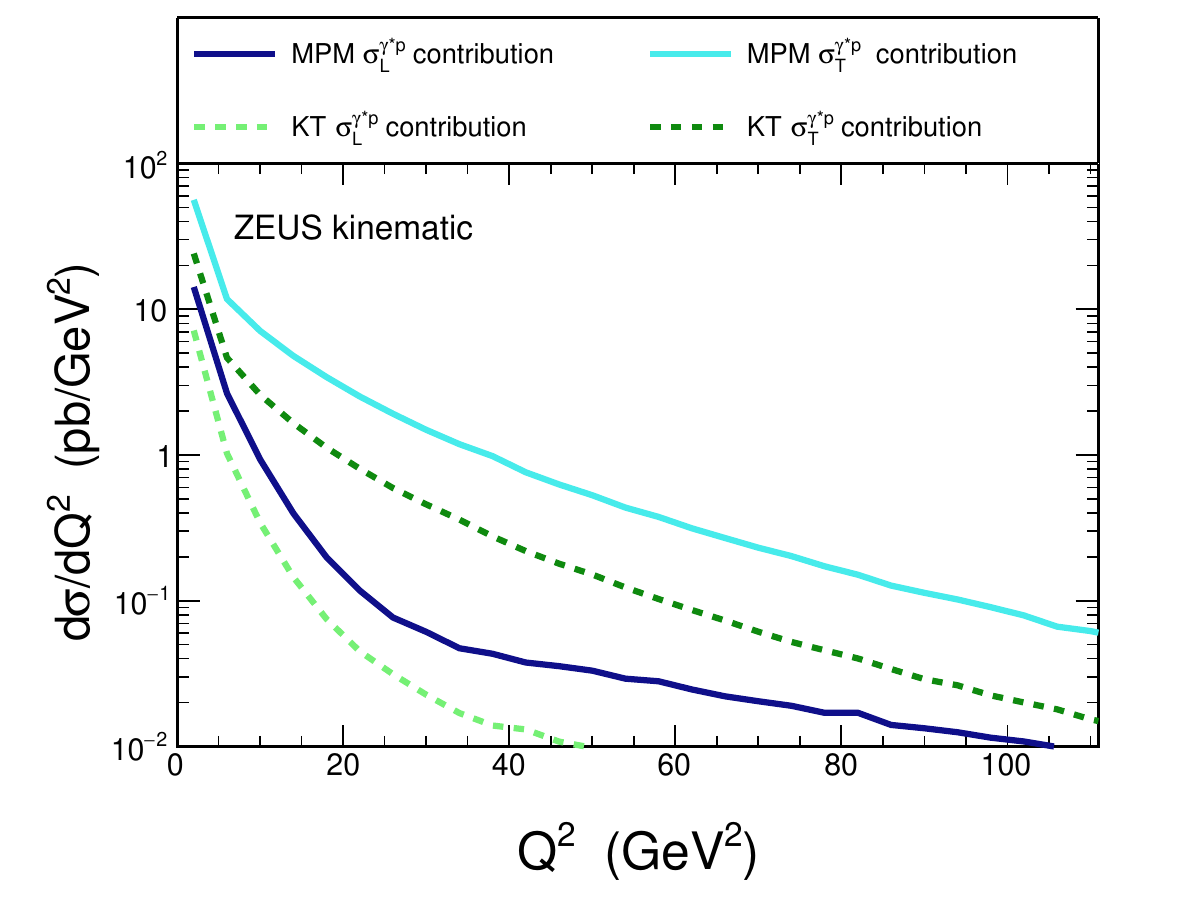}
  \caption{$Q^{2}$ distributions of transverse and longitudinal components of the cross section for the MPM and the KT GTMDs for H1 (left) and ZEUS (right) kinematics without $Q^{2}$ cuts.}
 \label{dsig_dQ2_TL}
\end{figure}

\begin{figure}
  \centering 
  \includegraphics[width=0.47\textwidth]{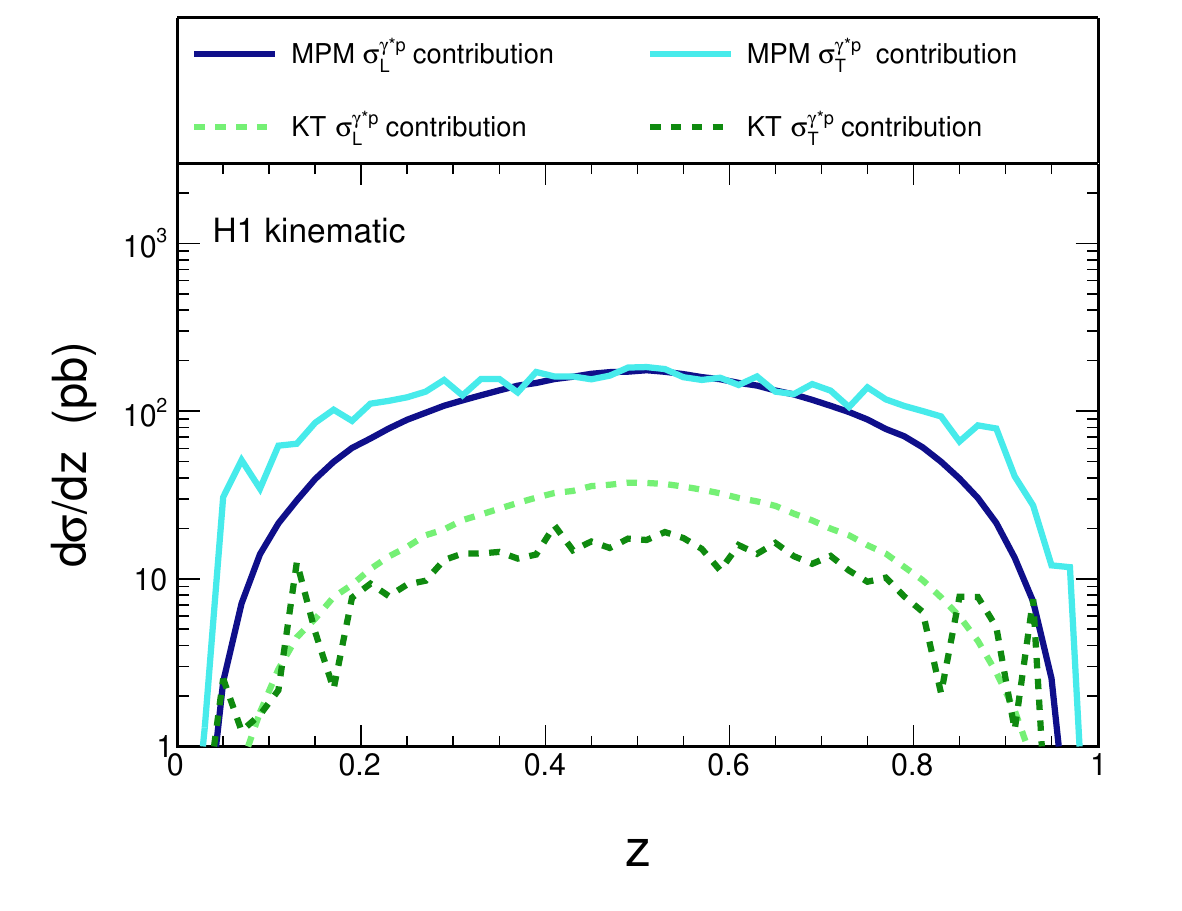}
  \includegraphics[width=0.47\textwidth]{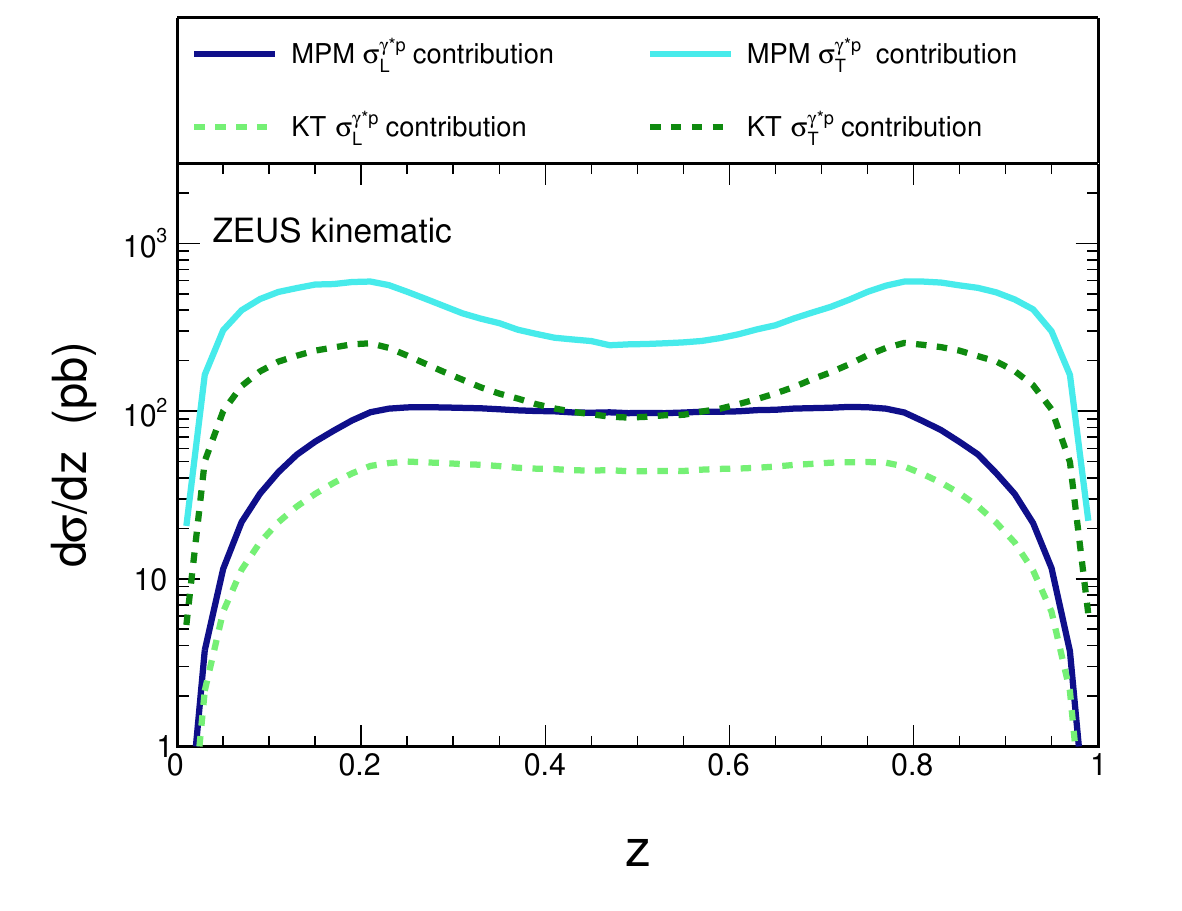}
  \caption{$z$ distributions of transverse and longitudinal components of the cross section for the MPM and the KT GTMDs for H1 (left) and ZEUS (right) kinematics.}
 \label{dsig_dz_TL}
\end{figure}

Finally in Fig. \ref{dsig_dQ2_TL} and Fig. \ref{dsig_dz_TL} we wish to show separately the transverse and longitudinal components to the cross section as a function of photon virtuality and light-front momentum fraction. At high $Q^{2}$ the longitudinal component dominates only for the H1 kinematics, which is associated with their behavior in the $z$ distributions. The relative relation between contributions depends on the model of GTMD.

\begin{figure}
  \centering
  \includegraphics[width=0.47\textwidth]{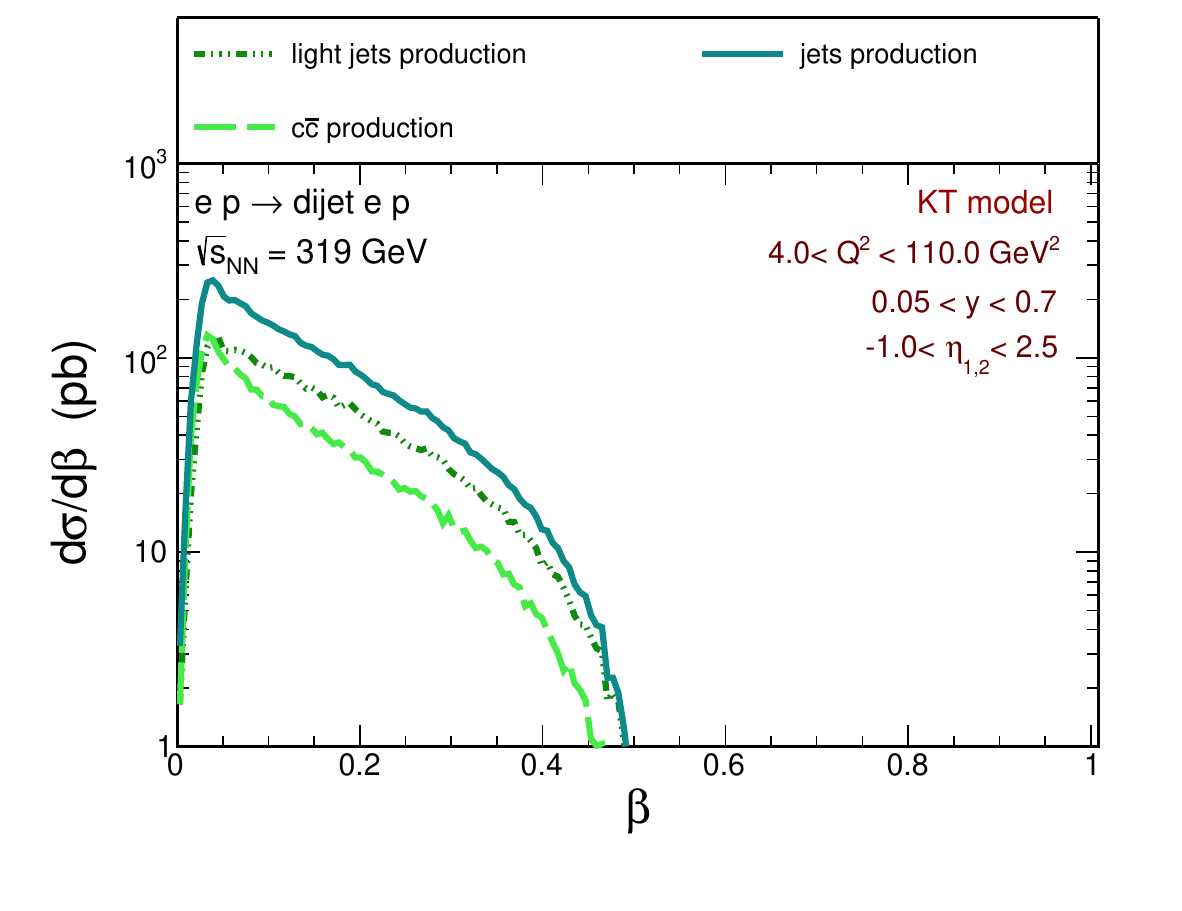}
  \includegraphics[width=0.47\textwidth]{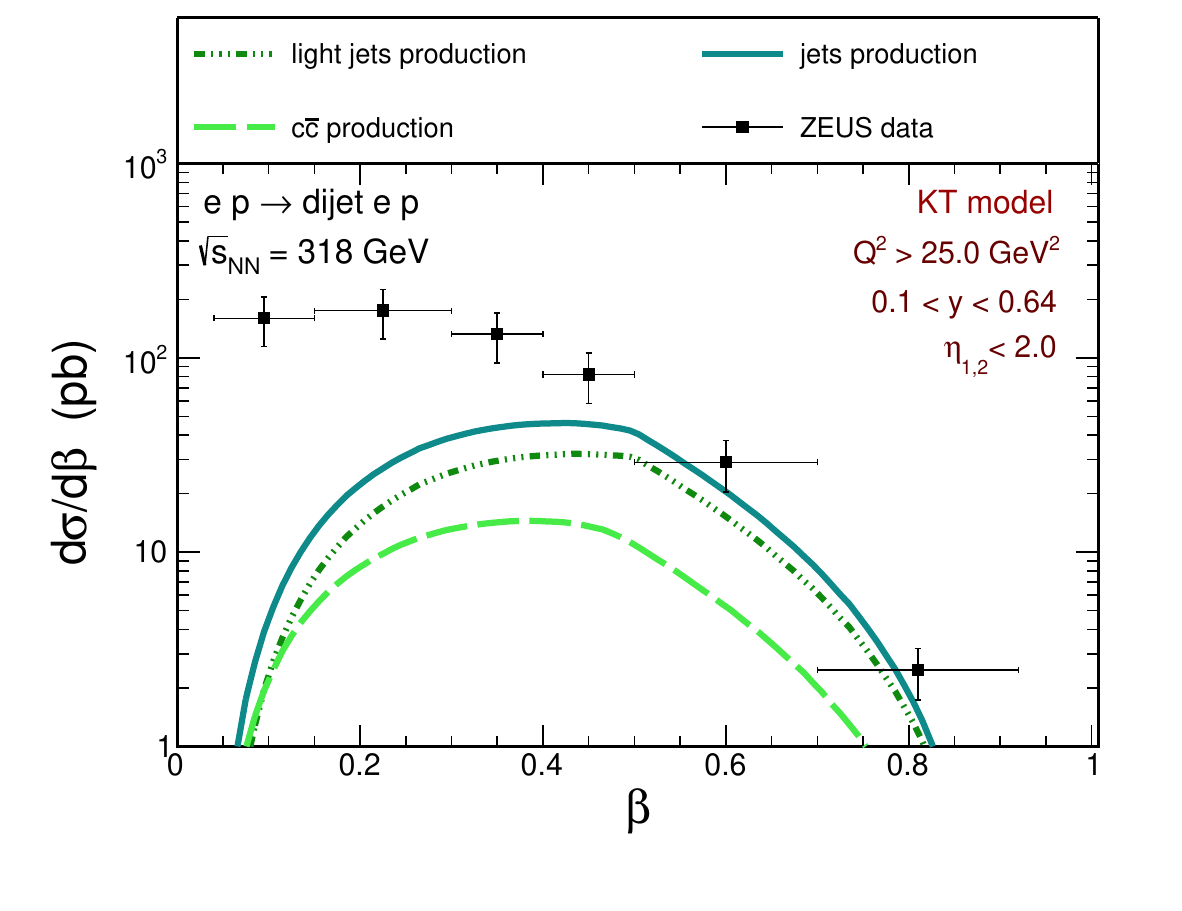}
  \caption{Contributions of light and $c\bar{c}$ dijets as a function of $\beta$ for the KT GTMD for H1 (left) and ZEUS (right) kinematics.}
\label{dsig_dbetacz}
\end{figure}

\begin{figure}
  \centering
  \includegraphics[width=0.47\textwidth]{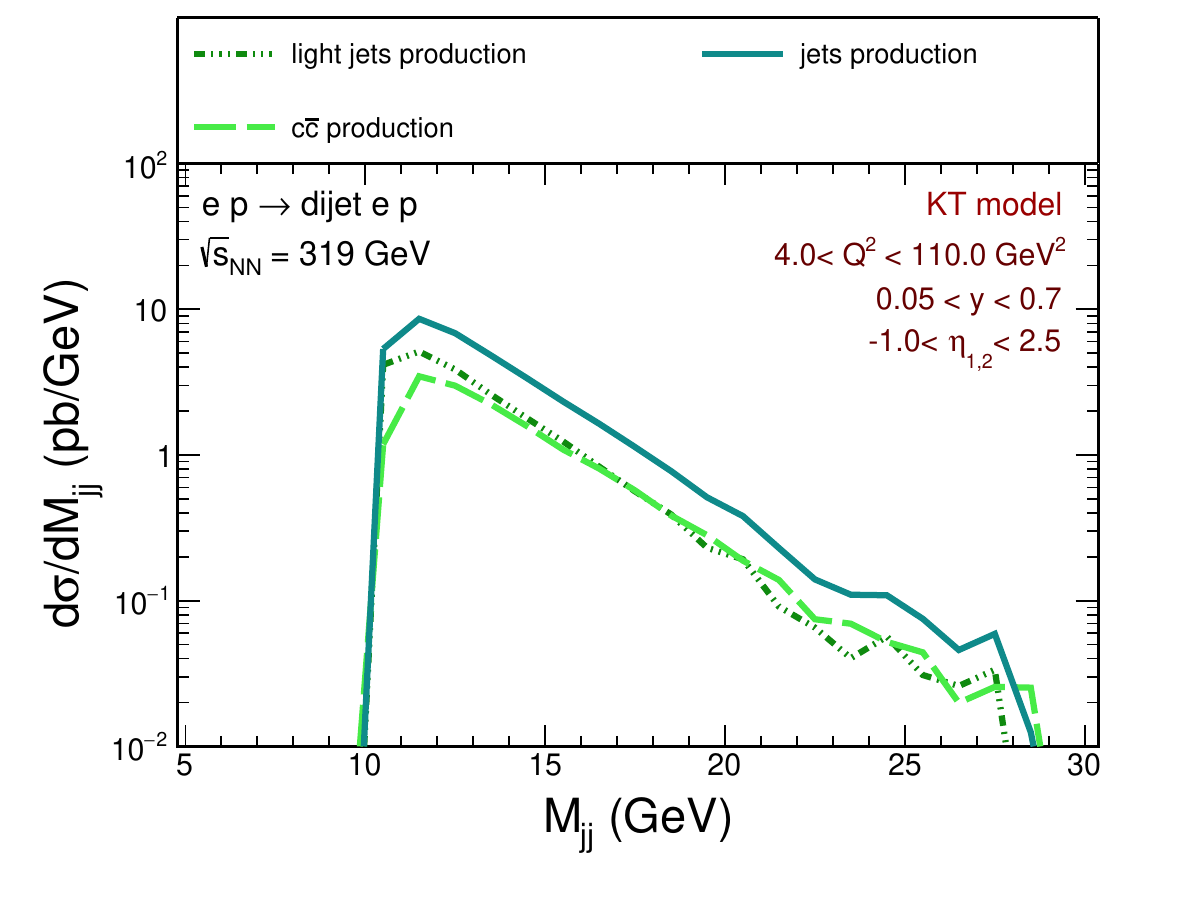}
  \includegraphics[width=0.47\textwidth]{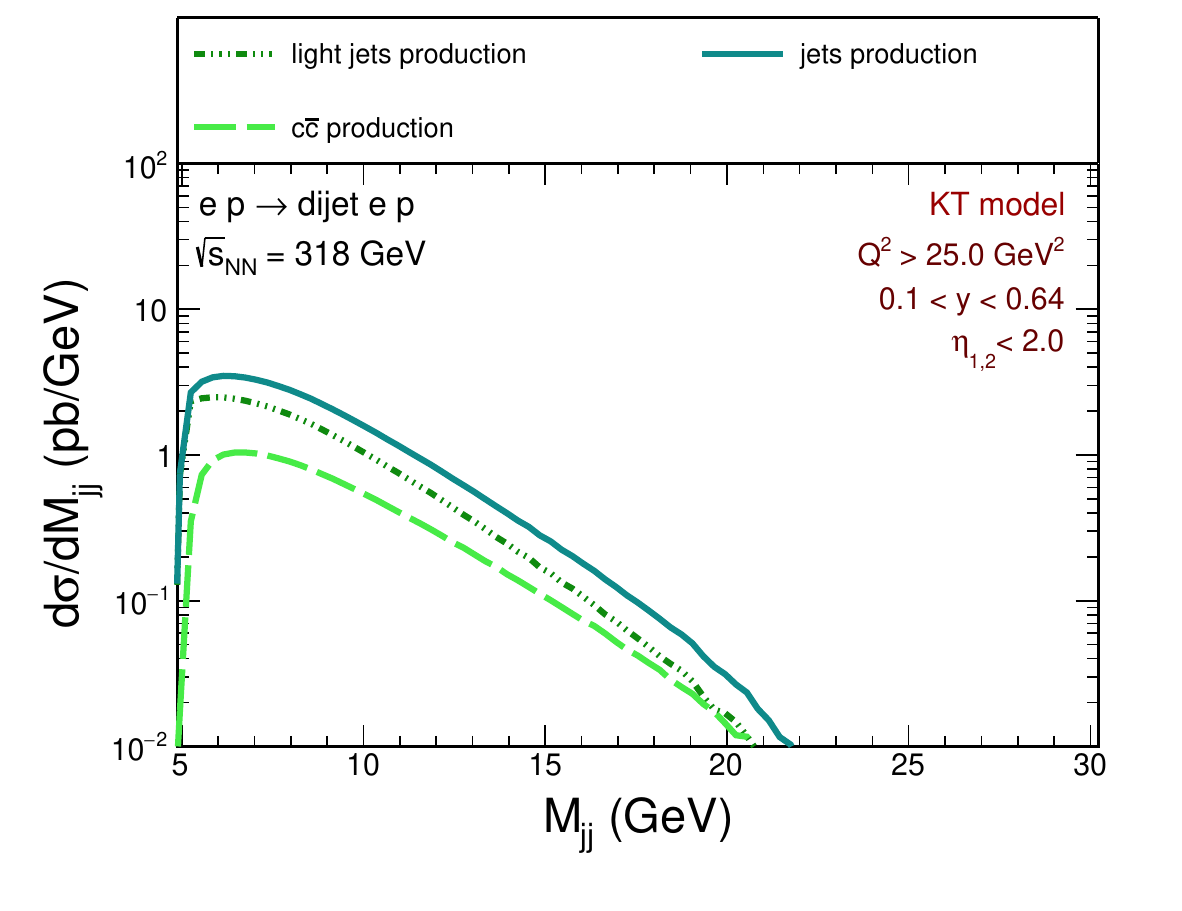}
  \caption{Contributions of light and $c\bar{c}$ dijets as a function of $M_{jj}$ for the KT GTMD for H1 (left) and ZEUS (right) kinematics.}
\label{dsig_dminvcz}
\end{figure}

\begin{figure}
  \centering
  \includegraphics[width=0.47\textwidth]{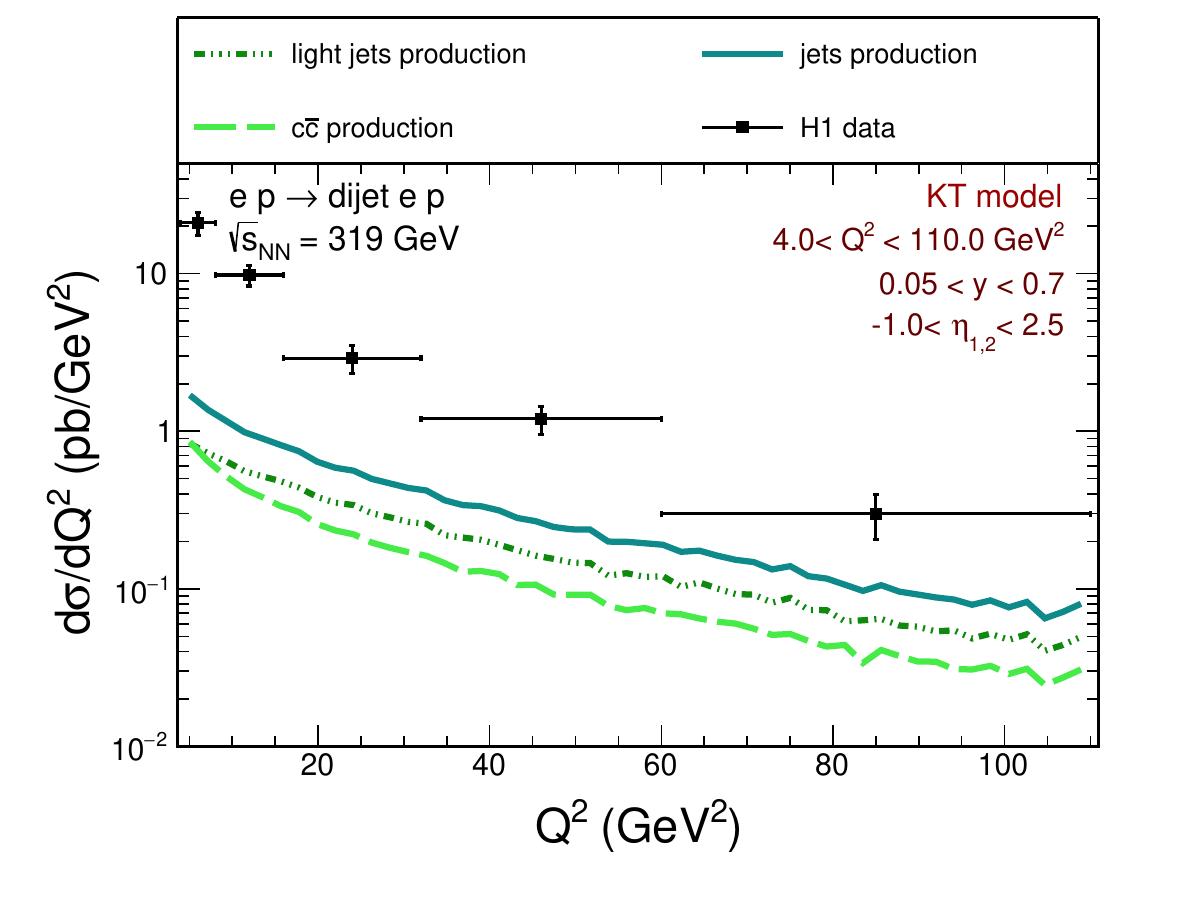}
  \includegraphics[width=0.47\textwidth]{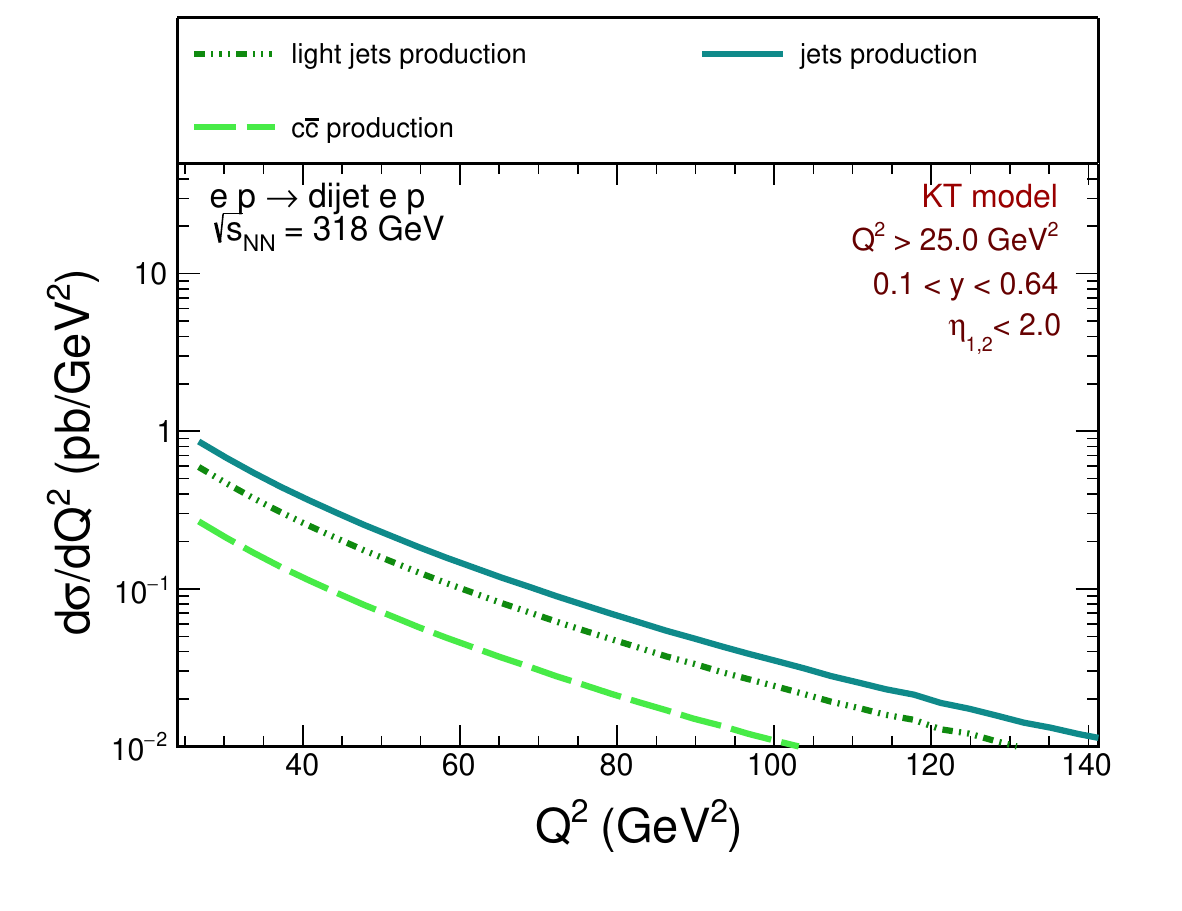}
  \caption{Contributions of light and $c\bar{c}$ dijets as a function of $Q^{2}$ for the KT GTMD for H1 (left) and ZEUS (right) kinematics.}
\label{dsig_dQ2cz}
\end{figure}

Up to now we include only light-quark jets similarly to ref. \cite{Boer:2021upt}. In Fig.~\ref{dsig_dbetacz}, ~\ref{dsig_dminvcz}, ~\ref{dsig_dQ2cz} we show what happens when we add in addition $c\bar{c}$ dijets. Including $c\bar{c}$ may cause increase of the cross section by as much as $50\%$ for ZEUS and $70\%$ for H1 kinematics. This goes beyond the scope of the present analysis.

Another topic which requires further studies is inclusion of Sudakov effects (see e.g. \cite{Hatta2020}), but this goes beyond the scope of the present paper.

\begin{figure}
  \centering
  \includegraphics[width=0.48\textwidth]{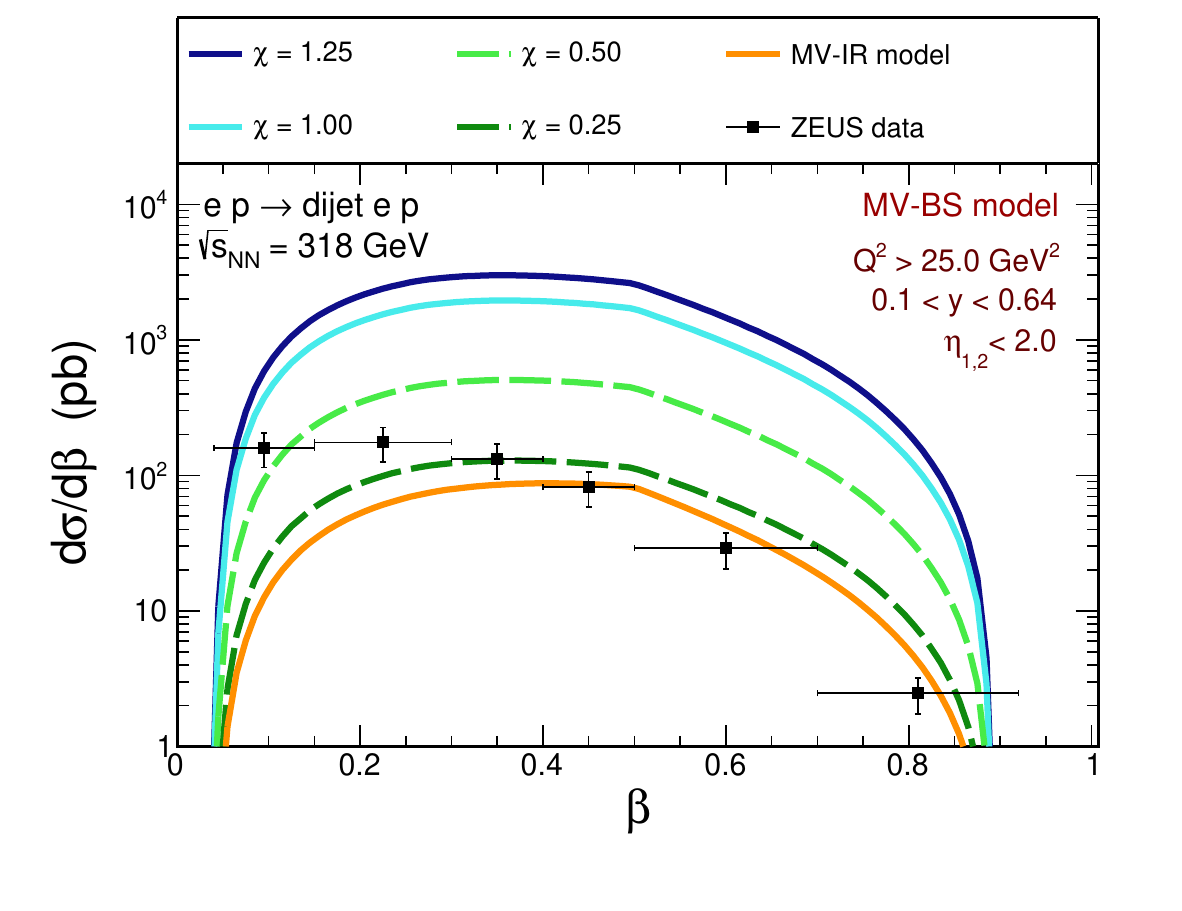}
  \includegraphics[width=0.48\textwidth]{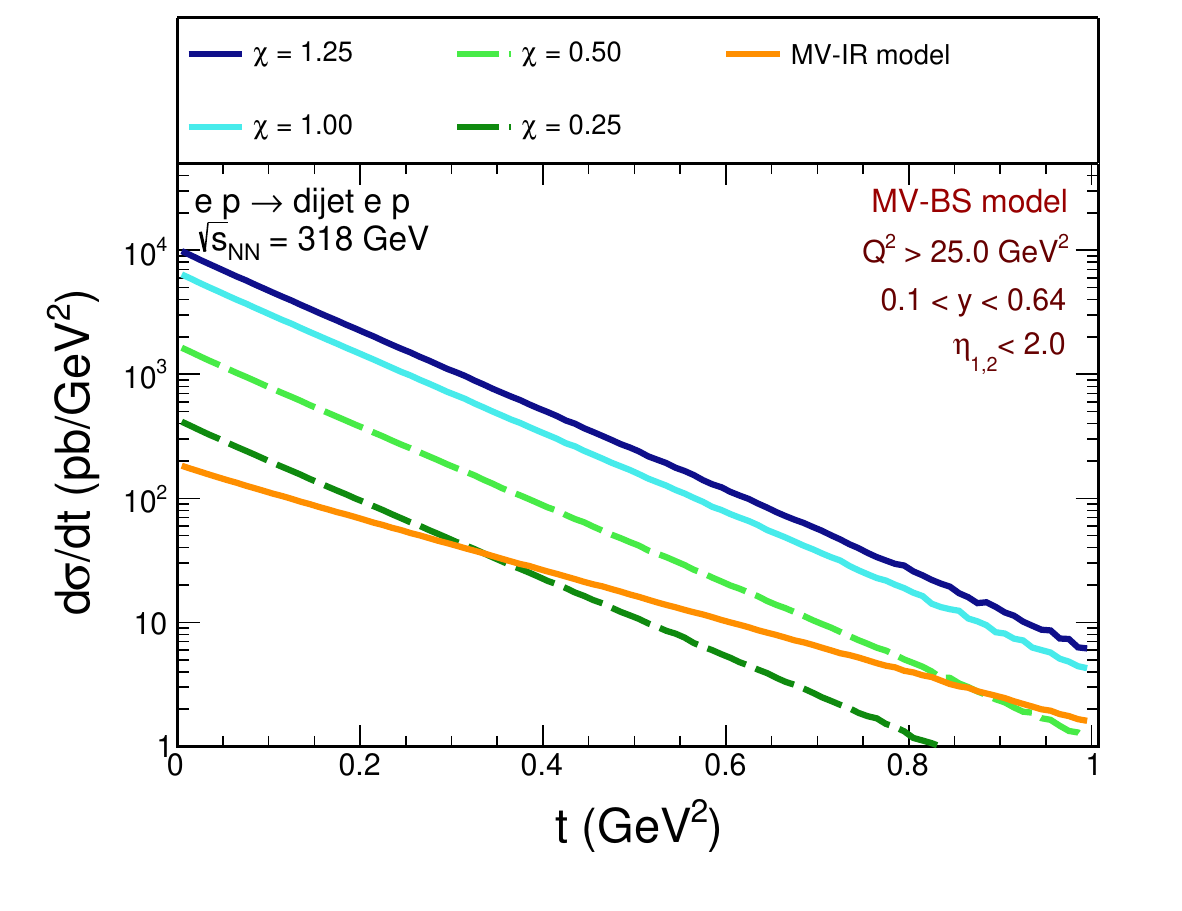}
  \caption{$\beta$ and $t$ distributions for ZEUS kinematic for the MV-BS model with different $\chi$ values.}
\label{dsig_dchiz}
\end{figure}

As shown above, the MV-BS model overtimes some distributions, therefore we also show distributions in $\beta$ and Mandelstam $t$ of the MV-BS model taking into account four different values of the $\chi$ factor in Fig. ~\ref{dsig_dchiz}. It is shown that its value have no impact to the shapes of distributions but changes their normalization of the cross section. 

\section{Conclusions}
\label{sec:conclusions}
In the present paper we have discussed dijet correlations in the $e p \to e p j j$ process. The corresponding differential distributions have been calculated using gluon GTMDs (generalized transverse momentum dependent gluon distributions) of the proton. Different GTMDs from the literature have been employed.  We have calculated distributions in many kinematical variables and compared our results to existing H1 and ZEUS data. The Boer-Setyadi MPM and MV-IR GTMDs reasonably well describe many observables for the H1 kinematics, but fail to describe distributions in $x_{\Pom}$ distribution and strongly overshoots the cross section differential in $\beta$ for the ZEUS kinematics. 
Here it needs to be pointed out that it is the ZEUS kinematics, where gluon GTMDs are expected to contribute. 
Some of the other GTMD distributions are consistent with the H1 and ZEUS data. 
The most realistic gluon distributions appear to be the KT and GBW GTMDs, which give a very small contribution for H1, as it should be, and give a sizeable contribution at $\beta > 0.5$ for the ZEUS cuts.
This means that the considered mechanism is not sufficient.
In view of the analysis presented in \cite{ZEUS:2016}, one should include also the diffractive excitation of the $q \bar q g$ component. Alternatively, one could try to explain the data within the resolved pomeron picture. This will be done elsewhere.

We have also calculated azimuthal angle correlations between the sum and difference of jet transverse momenta. As we use forward impact factors and our GTMDs have no elliptic part, these correlations are purely an effect of experimental cuts. As we have demonstrated, the cuts on jet transverse momenta generate modulation in $\phi_{\Delta_{\perp}P_{\perp}}$ which could be misidentified as the presence of elliptic gluon distribution. 

In Ref. \cite{PhysRevD.100.074020} only the GBW distribution was used for the dijet production, and the results shown in this paper are similar to our calculations for the same model of GTMD. Our analysis includes also a comparison of several different GTMD models and presents the distributions in many more kinematic variables.

\bibliography{main}
\end{document}